\documentclass[fleqn,usenatbib]{mnras}

\usepackage{times}

\usepackage[T1]{fontenc}
\usepackage{ae,aecompl}
\usepackage{hyperref}
\usepackage[utf8]{inputenc}
\usepackage{wrapfig,lipsum,booktabs}
\usepackage{longtable}
\usepackage{float}
\usepackage{rotating}
\usepackage{lscape}
\usepackage{tabularx}    
\usepackage{siunitx}     
\usepackage{adjustbox}
\usepackage{rotating}

\usepackage{color}

\DeclareRobustCommand{\VAN}[3]{#2}
\let\VANthebibliography\thebibliography
\def\thebibliography{\DeclareRobustCommand{\VAN}[3]{##3}\VANthebibliography}

\usepackage{cuted}
\usepackage{natbib}
\usepackage{graphicx}	
\usepackage{amssymb}	
\usepackage{amsmath} 
\usepackage{multirow}
\usepackage{epstopdf}
\DeclareUnicodeCharacter{2014}{\dash}

\title[The spin of the black hole in XTE~J1859+226]{Measuring Black Hole Spins through X-ray Reflection Spectroscopy and the Relativistic Precession Model: the case of XTE~J1859+226}

\author[Mall et al.]{
Gitika~Mall,$^{1}$ 
Honghui Liu,$^{1}$
Cosimo~Bambi,$^{1,2}$\thanks{Email: \href{mailto:bambi@fudan.edu.cn}{bambi@fudan.edu.cn}}
James F. Steiner,$^{3}$ and
\newauthor
Javier A. Garc{\'\i}a $^{4,5}$
\\
$^{1}$Center for Field Theory and Particle Physics and Department of Physics, Fudan University, 200438 Shanghai, China\\
$^{2}$School of Natural Sciences and Humanities, New Uzbekistan University, Tashkent 100007, Uzbekistan\\
$^{3}$Harvard-Smithsonian Center for Astrophysics, Cambridge, MA 02138, USA\\
$^{4}$Cahill Center for Astronomy and Astrophysics, California Institute of Technology, Pasadena, CA 91125, USA\\
$^{5}$Dr. Karl Remeis-Observatory and Erlangen Centre for Astroparticle Physics, D-96049 Bamberg, Germany\\
}

\pubyear{2021}

\begin{document}
\label{firstpage}
\pagerange{\pageref{firstpage}--\pageref{lastpage}}
\maketitle

\begin{abstract}
The development of techniques to measure accurately black hole spins is crucial to study the physics and astrophysics of these objects. X-ray reflection spectroscopy is currently the most popular method to estimate the spins of accreting black holes; so far it has provided a spin measurement of about 40 stellar-mass black holes in X-ray binaries and 40 supermassive black holes in active galactic nuclei. The relativistic precession model (RPM) is another method to measure the spins of stellar-mass black holes: it requires the measurement of the frequencies of three simultaneous quasi-periodic oscillations and can potentially provide precise estimates of the black hole mass and spin. However, the two methods do not seem to provide consistent results when applied to the same sources, which questions the reliability and accuracy of these measurements. Recently, the RPM has been applied to infer the spin of the black hole in XTE~J1859+226. The authors found $a_* = 0.149 \pm 0.005$ (68\% CL). There are no other spin measurements of this source. We looked for archived \textsl{RXTE} observations of XTE~J1859+226 with blurred reflection features and found 23~spectra suitable for measuring the spin. We employed two different models with {\tt relxill} and {\tt relxillD} and obtained a higher spin value from all these fits. From simultaneous fitting of 7~spectra of higher quality, we found $a_* = 0.986^{+0.001}_{-0.004}$ and $a_* =0.987 \pm 0.003$ (90\% CL, statistical) with {\tt relxill} and {\tt relxillD}, respectively. Our results confirm the discrepancy between the spin measurements inferred from the two techniques.

\end{abstract}




\section{Introduction}
\label{sec:Intro}

Einstein's theory of general relativity predicts the existence of black holes (BHs). These evolutionary remnants of massive stars can be fully described by three parameters: mass, spin (angular momentum), and electric charge~\citep[see, e.g.,][]{2012LRR....15....7C}. Since astrophysical BHs are believed to have negligible electric charge~\citep{2017bhlt.book.....B}, they should be completely specified by the values of their mass and spin. Placing constraints on the BH masses and spins is crucial to study physics and the astrophysics of these objects~\citep{2014SSRv..183..277R,2021SSRv..217...65B}. Spin measurements are challenging because they require observations that probe the immediate environment of the BH event horizon.

In the case of stellar-mass BHs, it is usually thought that the value of their spin is natal~\citep{10.1046/j.1365-8711.1999.02482.x}. If the BH is in a low-mass X-ray binary system (LMXB), the companion star is up to about one Solar mass, and even if the BH swallows the whole star it cannot appreciably change the value of its spin because its mass is about ten times higher. If the BH is in a high-mass X-ray binary system, the companion star is a blue giant and its lifetime is too short to transfer a significant amount of mass to the BH and change its spin even if the BH accretes at its Eddington limit. If this picture is correct, the spins of stellar-mass BHs should be explained by the mechanism of gravitational collapse of the progenitor stars \citep[but see][for different conclusions]{2015ApJ...800...17F}.

Today there are three leading techniques for measuring BH spins: the continuum-fitting method (CFM)~\citep{1997ApJ...482L.155Z,2011CQGra..28k4009M,2014SSRv..183..295M}, X-ray reflection spectroscopy (XRS)~\citep{2006ApJ...652.1028B,2014SSRv..183..277R,2021SSRv..217...65B}, and the analysis of the gravitational waves (GWs) from the coalescence of binary BHs~\citep{PhysRevLett.112.251101}. CFM and XRS involve X-ray observations of accreting BHs with geometrically thin and optically thick disks, while GWs can be used for compact binary systems BH-BH (or BH-neutron star) during the merger of the two bodies.

In the CFM, we fit the thermal spectrum of geometrically thin and optically thick accretion disks. The standard framework is the Novikov-Thorne disk model \citep{1973blho.conf..343N,1974ApJ...191..499P}. Assuming the Kerr spacetime and that the inner edge of the accretion disk is located at the innermost stable circular orbit (ISCO), the model has five free parameters: the BH mass, the BH spin, the BH mass accretion rate, the BH distance, and the inclination angle of the disk with respect to the line of sight of the observer. If we can have independent estimates of the BH mass, distance, and inclination angle of the disk, for example from optical observations, we can fit the data and infer the BH spin and the BH mass accretion rate. Since the thermal spectrum of the disk is peaked in the soft X-ray band for stellar-mass BHs and in the UV band for supermassive BHs, this technique is normally used only to measure the spins of stellar-mass BHs. In the case of supermassive BHs, dust absorption in the UV band inhibits an accurate measurement of the thermal spectra. As of now, this technique has provided an estimate of the spin of about 20~stellar-mass BHs~\citep{2023ApJ...946...19D}.

XRS is the analysis of the disk's reflection spectrum, which is generated when an X-ray source illuminates the disk. The reflection spectrum in the rest-frame of the gas in the disk is normally characterized by narrow fluorescent emission lines in the soft X-ray band and by a Compton hump with a peak around 20-30~keV~\citep{2005MNRAS.358..211R,2010ApJ...718..695G}. The reflection spectrum of the whole disk as seen by a distant observer appears blurred because of relativistic effects in the strong gravity region~\citep{1989MNRAS.238..729F,1991ApJ...376...90L,2017bhlt.book.....B}. This technique can be used for both stellar-mass BHs in XRBs and supermassive BHs in active galactic nuclei, and it is currently the most widely adopted method to estimate the spins of supermassive BHs. So far XRS has provided a spin measurement for about 40~stellar-mass BHs and for a similar number of supermassive BHs~\citep{2021SSRv..217...65B,2023ApJ...946...19D}.

With current observational facilities, GWs can only be used to measure the spins of stellar-mass BHs in compact binary systems. GW signals are analysed to estimate the posterior probability distribution for the binary parameter values~\citep{2016PhRvL.116x1102A}. The spin parameter value that is well constrained using GW observations is the vector combination of the two spins, $\chi$. The spins themselves are poorly constrained. The GW approach has an upper hand over electromagnetic observational techniques as the latter could be affected by the uncertainties of the astrophysical model but the former is limited due to the low signal-to-noise ratio (SNR) and systematics associated with the GW modeling. It is notable that X-ray techniques, and in particular XRS, often provide high values for the dimensionless spin parameters $a_*$, close to the upper limit 1~\citep{2021SSRv..217...65B,2023ApJ...946...19D,2023ApJ...950....5L,2022MNRAS.512.2082L}. On the contrary, with GWs, we normally find low values of the BH spin parameters~\citep[see, e.g.,][]{2018ApJ...868..140T}.

Quasi-periodic oscillations (QPOs) are relatively narrow peaks at characteristic frequencies in the X-ray power density spectra of neutron stars and BHs~\citep[see, e.g.,][]{2004astro.ph.10551V,2006ARA&A..44...49R}. The exact physical mechanism responsible for the production of these QPOs is not yet well understood. The centroid frequency measurement of high frequency ($\sim$ 100 Hz) QPOs can be highly precise and be associated with matter orbiting near the ISCO. In the relativistic precession model (RPM), the QPO frequencies are associated with the three fundamental frequencies of equatorial circular orbits of a test-particle~\citep{1998ApJ...492L..59S,1999ApJ...524L..63S,1999PhRvL..82...17S}. In the Kerr spacetime, these frequencies depend on three parameters: the BH mass, the BH spin parameter, and the radial coordinate of the oscillation. \citealt{2014MNRAS.437.2554M} tested the RPM where Type-C QPOs \citep{2004A&A...426..587C} found in LMXBs are associated with the nodal precession frequency (or Lense–Thirring frequency), while the periastron precession frequency and the orbital frequency correspond, respectively, to the lower and upper high-frequency QPOs. If we observe these three frequencies simultaneously, we can infer the values of the BH mass, the BH spin, and the radial coordinate of the oscillation. Since it is relatively easy to get a very precise measurement of the QPO frequencies, if this interpretation is correct we can obtain very precise measurements of the BH mass and BH spin.

The RPM has so far provided a spin measurement for 5 stellar-mass BHs (see Table~\ref{tab:table1}). The first spin measurement through the RPM was reported in \citet{2014MNRAS.437.2554M} for the BH in GRO~J1655--40. From the simultaneous observation of a type-C QPO and two high-frequencies QPOs, the authors inferred the BH spin parameter $a_* = 0.290 \pm 0.003$ (68\% CL), which is significantly lower than the values inferred from the CFM and XRS. However, the BH spin measurement reported in \citet{2009ApJ...697..900M} from the analysis of the reflection spectrum was obtained in the early ages of XRS and GRO~J1655--40 is a complicated source with strong absorption (see also \citealt{2009MNRAS.395.1257R}). The RPM was then used to infer the BH spin of XTE~J1550-564 in \citet{2014MNRAS.439L..65M}; however, for this source, we do not have any observation with three simultaneous QPOs, so the authors determined the BH spin combining the detection of two simultaneous high-frequency QPOs with the dynamical measurement of the BH mass reported in \citet{2011ApJ...730...75O}. The reported spin measurement $a_* = 0.34 \pm 0.01$ (68\% CL) is consistent with the spin measurement from the CFM and XRS, but the measurements from CFM and XRS have large uncertainties. The RPM was again used to measure the spin of the BH in MAXI~J1820+070 in \citet{2021MNRAS.508.3104B}. The authors found $a_* = 0.799_{-0.015}^{+0.016}$ (68\% CL), which is lower than the measurement obtained from XRS in \citet{2023ApJ...946...19D} \footnote{The very low spin measurement from the CFM reported in \citet{2021ApJ...916..108Z} is probably because the authors analyzed observations in which the disk was truncated.}. However, in this case, the authors did not observe a QPO triplet instead they observed a low-frequency QPO and broad noise features in the \textsl{NICER} observations.
\cite{motta2023rethinking} re-analysed \textsl{RXTE} observations of GRS 1915+105 and identified the 67~Hz QPO as the Lense-Thirring frequency at the ISCO yielding a value of $0.706 \pm 0.034$ (68\% CL) for the black hole spin. The measurements from CFM and XRS provide a spin measurement very close to 1 (see \citealt{McClintock_2006} for the CFM and \citealt{Miller_2013} for XRS). 
Lastly, \citet{2022MNRAS.517.1469M} have recently used the RPM to measure the spin of the BH in XTE~J1859+226. Their measurement is $a_* = 0.149 \pm 0.005$ (68\% CL) and, as of now, there are no spin measurements of this source from CFM and XRS. It is worth noting that in the case of XTE~J1859+226, the RPM did return the correct values for the mass of the BH which was consistent with the value obtained from optical measurements. In the case of GRO~J1655-40, the value for the mass of the BH returned by RPM is consistent with one of the measurements obtained using the optical band (\citealt{2002MNRAS.331..351B}), however, there are other measurements as well (e.g. \citealt{1997ApJ...477..876O}).

\begin{table*}
\centering
		\caption{BHs with a spin measurement from the relativistic precession model reported in the literature. The table also shows the existing spin measurements from the continuum-fitting method and X-ray reflection spectroscopy of these objects. $^\dag$ In \citet{2014MNRAS.439L..65M}, the BH spin is obtained within the relativistic precession model from the detection of two simultaneous QPOs and the dynamical measurement of the BH mass.}
{\scriptsize
  \renewcommand{\arraystretch}{1.4}
	\begin{tabular}{c c c c c c c}
	    \hline
		\hline
        \multirow{2}{2em}{Source} & \multicolumn{6}{c}{Spin measurements}\\\cline{2-7}
		 & CFM & CL & XRS & CL & RPM & CL \\\hline
  \hline
GRO J1655--40 & $\sim$ 0.65–0.75
\citep{2006ApJ...636L.113S} & 90\% & $0.98 \pm 0.01$ \citep{2009ApJ...697..900M} & 68\% & $0.290 \pm 0.003$ \citep{2014MNRAS.437.2554M} & 68\% \\
XTE J1550--564 & $0.34^{+0.37}_{-0.45}$ \citep{2011MNRAS.416..941S} & 90\% & $0.55^{+0.15}_{-0.22}$ \citep{2011MNRAS.416..941S} & 90\% & $0.34 \pm 0.01$ \citep{2014MNRAS.439L..65M}$^\dag$ & 68\% \\
MAXI J1820+070 & $0.14^{+0.09}_{-0.09}$ \citep{2021ApJ...916..108Z} & 68\% & $0.988^{+0.006}_{-0.028}$ \citep{2023ApJ...946...19D} & 68\% & $0.799^{+0.016}_{-0.015}$ \citep{2021MNRAS.508.3104B} & 68\%\\
GRS 1915+105 & $>$ 0.98 \citep{McClintock_2006} & 99\% & $0.98 \pm 0.01$ \citep{Miller_2013} & 68\% & $0.706 \pm 0.034$ \citep{motta2023rethinking} & 68\% \\
XTE J1859+226 & -- & & -- &  & $0.149 \pm 0.005$ \citep{2022MNRAS.517.1469M} & 68\%\\
  \hline
		\hline
	\end{tabular}
	}
	\label{tab:table1}
 \end{table*} 

The goal of this work is to measure the spin of the BH in XTE~J1859+226 through XRS in order to compare such a measurement with that inferred from the RPM in \citet{2022MNRAS.517.1469M}. To do this, we looked for spectra with relativistic reflection features among the 138~\textsl{RXTE} pointings of XTE~J1859+226. We selected 23 distinct observations with the strongest relativistic features and we analyzed these spectra with the reflection models {\tt relxill} and {\tt relxillD}~\citep{2013MNRAS.430.1694D,2014ApJ...782...76G}. After performing the spectral analysis, we found the BH spin measurement $a_*$ to be $0.986^{+0.001}_{-0.004}$ and $0.987 \pm 0.003$ (90\% CL, statistical), respectively, which is clearly inconsistent with the RPM spin estimate.

The content of our paper is as follows. In Section~\ref{1.5}, we briefly introduce our source, XTE~J1859+226. In Section~\ref{2}, we describe the data reduction process. In Section~\ref{sec:3}, we present our spectral analysis of the source. In Section~\ref{sec:4}, we discuss our results.


\section{XTE~J1859+226}
\label{1.5}

XTE J1859+226 was first detected as an X-ray transient source by the All–Sky–Monitor (ASM) aboard the Rossi X-Ray Timing Explorer (\textsl{RXTE}) \citep{1993A&AS...97..355B} on 9 October 1999 \citep{1999IAUC.7274....1W}. A follow-up observation of the source with the \textsl{RXTE}/PCA (Proportional Counter Array, 2–60 keV) revealed a hard power-law dominated spectrum \citep{1999IAUC.7274....2M}. Located at a distance of $\sim$~11~kpc \citep{2002MNRAS.334..999Z}, the transient source was also detected in radio \citep{1999IAUC.7278....1P} as well as in gamma-rays \citep{1999IAUC.7282....3M,1999IAUC.7291....2D}. Radio–X-ray correlations were observed \citep{2002MNRAS.331..765B}. Spectral and temporal characteristics also confirmed that the source was a BH candidate \citep{2001ApSSS.276..209M}. Optical photometry and spectroscopy observations revealed an orbital period of $6.58 \pm 0.05$~hr, a mass function $f(M) = 4.5 \pm 0.6$~$M_\odot$, and an upper limit of inclination of $i = 70$~deg \citep{2011MNRAS.413L..15C}. With the same procedure, these authors also determined the BH mass to be $M_{BH} = 7.7 \pm 1.3$~$M_\odot$. The measured optical extinction ${\rm E(B-V)} = 0.58 \pm 0.07$ \citep{2002MNRAS.331..169H} is consistent with the average $N_{\rm H}$ value of $2.21 \cdot 10^{21}$~cm$^{-2}$ derived from radio maps in the Dickey \& Lockman Survey\footnote{\hyperlink{http://heasarc.nasa.gov/cgi-bin/Tools/w3nh/w3nh.pl}{http://heasarc.nasa.gov/cgi-bin/Tools/w3nh/w3nh.pl}}. Both low-frequency ($\sim$~1--4~Hz and $\sim$~6~Hz) and high-frequency ($\sim$~150--187~Hz) QPOs have been reported \citep{2000ApJ...535L.123C,2000arxt.confE.104F} in the observations of XTE J1859+226 \citep{2004A&A...426..587C}.


\section{Data Reduction}
\label{2}

\begin{figure}
    \caption{The observation log of XTE J1859+229 by ASM on-board \textsl{RXTE}. Each red point represents the respective epoch number's average ASM unit count rate. The blue asterisk denotes the observations considered in this work.}
    \includegraphics[width=3.3in]{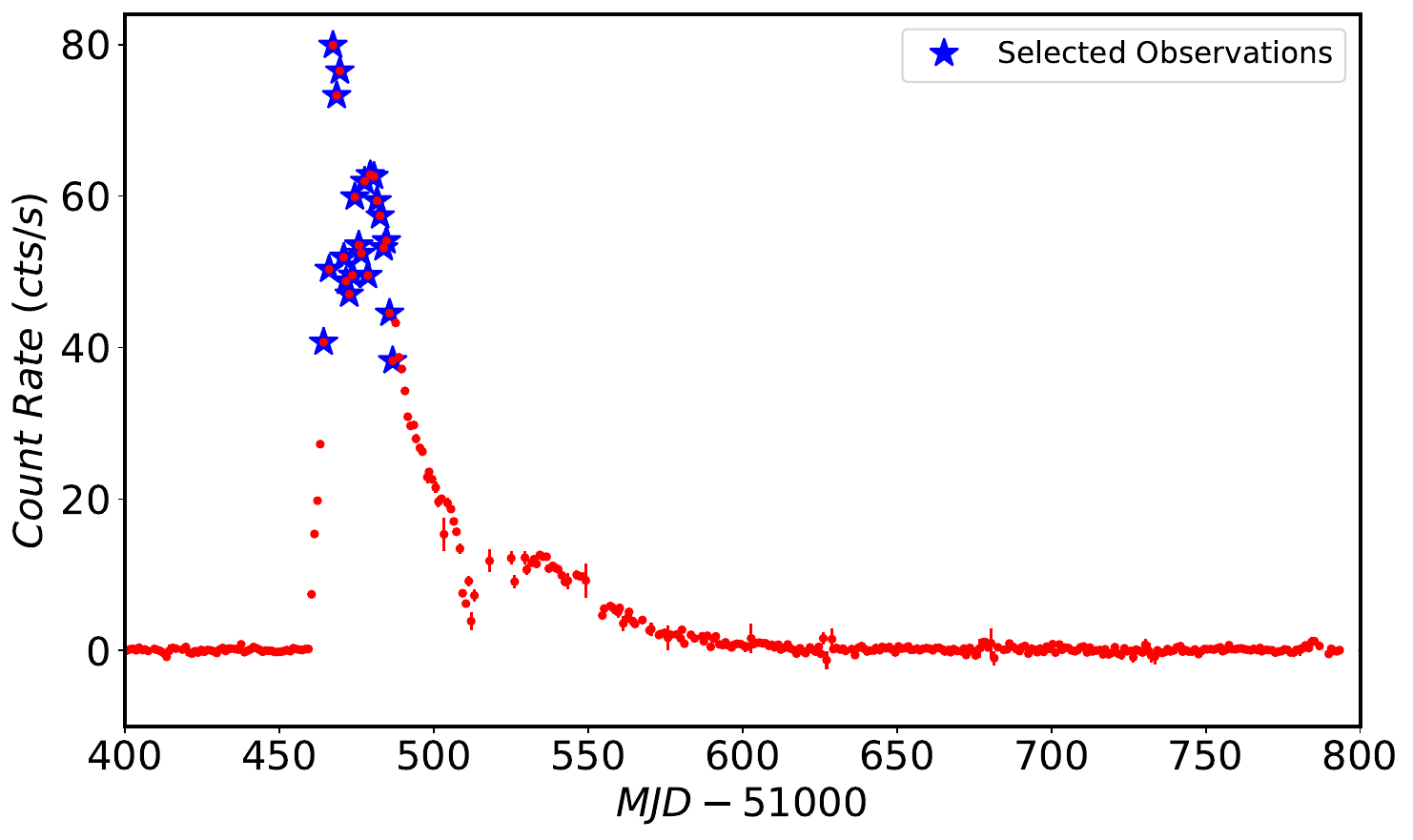}
    \label{fig:figure1a}
\end{figure}

\textsl{RXTE}/PCA extensively observed XTE J1859+226 during its outburst in the years 1999 and 2000 with 138 distinct pointings. The data are publically available in the \textsl{RXTE} archive via the HEASARC (High Energy Astrophysics Science Archive Research Center). We investigated all the available standard 2 data between MJD 51461 and 51749 (Figure~\ref{fig:figure1a}) by extracting the spectra from all five nearly identical Proportional Counting Units (PCUs) in the Proportional Counting Array (PCA) \citep{2006ApJS..163..401J} on-board the \textsl{RXTE} mission. We employed for our analysis only the spectra from PCU~2 as it is the most well-calibrated and has the most extensive coverage among all the units. We applied the publicly available tool \textit{pcacorr} for improving the calibration to a $\sim$~0.1\% systematic precision \citep{2014ApJ...794...73G}. The standard background model was applied and subtracted. We confine our spectral fitting in the 3–50~keV energy band.

\section{BROADBAND X-RAY SPECTRAL ANALYSIS}
\label{sec:3}
\begin{figure}
    \caption{ The Hardness-Intensity Diagram of the 1999 outburst of XTE J1859+226 as observed by \textsl{RXTE}. Each point represents data from one day of observation. The rainbow colour gradient represents the reflection strength. The observation points in the red box are used in this work. The green and blue stars represent the observations that were analysed in the sample sets G1 and G2, respectively.}
	\includegraphics[width=0.49\textwidth]{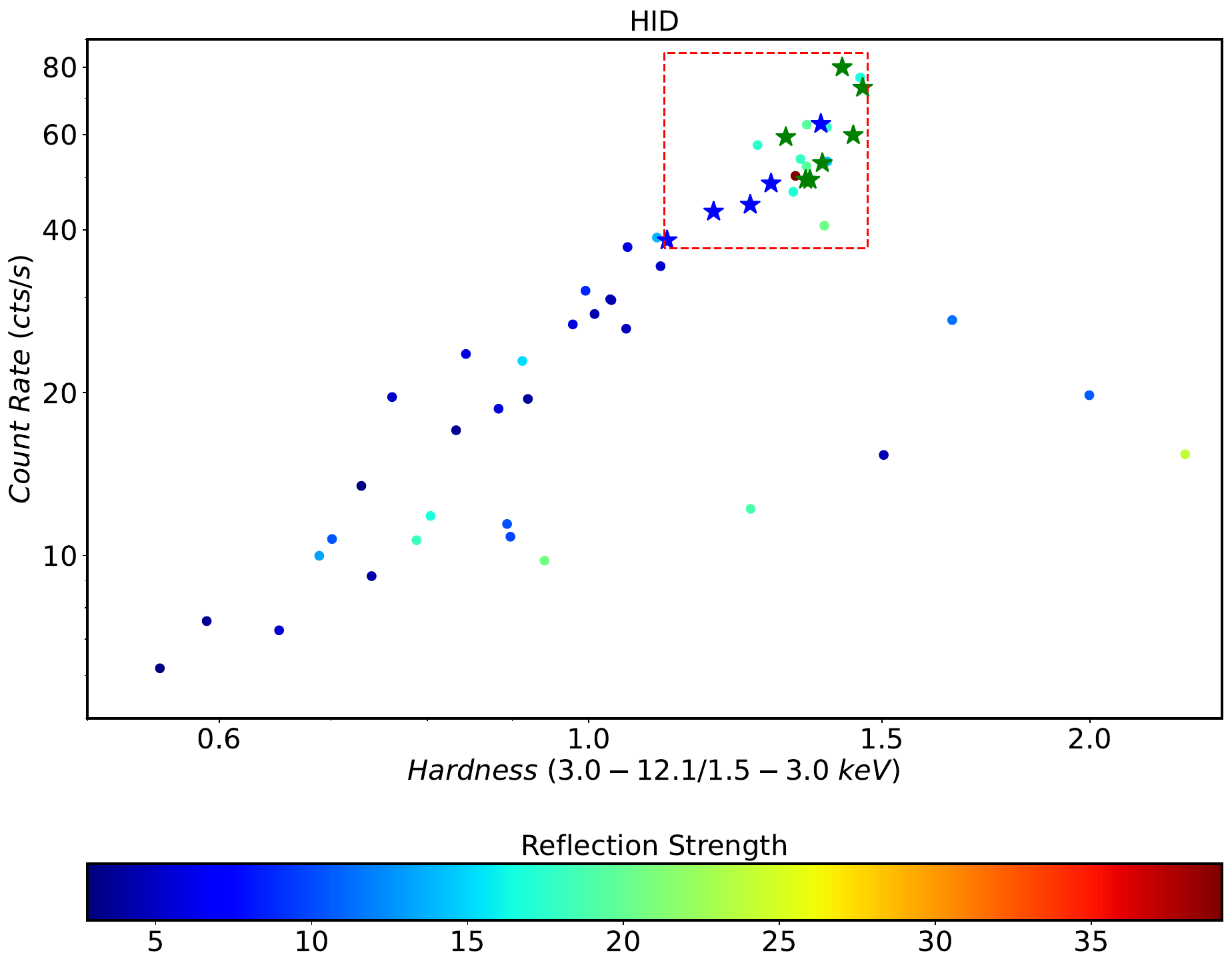}
    \label{fig:figure2}
\end{figure}

We used XSPEC v12.9.1 \citep{1996ASPC..101...17A} for the spectral analysis. To examine whether this ensemble of spectra obtained from distinct observations is suitable for our study, we initiated by fitting each of them with an absorbed power law model. We fixed the hydrogen column density in {\tt tbabs}, which calculates the cross section for X-ray absorption by the inter-stellar medium \citep{2000ApJ...542..914W}, to $N_{\rm H} = 0.2 \cdot 10^{22}$ atoms cm$^{-2}$ \citep{2014AdSpR..54.1678R}. Out of 138 spectra, 100 spectra had features resembling the relativistic reflection features. To quantify the strength of these features and select the best data for our analysis, we modeled every spectrum with a model consisting of a component to model the multi-color disc blackbody: {\tt diskbb}~\citep{1984PASJ...36..741M}, a component to model the continuum from the corona: {\tt nthcomp} and another component for the reflection features: {\tt relxillCp}~\citep{2013MNRAS.430.1694D,2014ApJ...782...76G}. In XSPEC notation, the model becomes: {\tt tbabs}*{(\tt diskbb} + {\tt nthcomp} + {\tt relxillCp}). To compute the flux contribution of every component separately, we then convoluted each of them with the model {\tt cflux} and calculated the total flux. The reflection component's strength was defined as the ratio of the flux of the reflection component to the total flux. For the selection process, we plotted the reflection strength of each observation on top of the hardness-intensity diagram where the intensity is one-day averaged counts in three energy bands viz. A-Band
(1.5-3.0 keV), B-Band (3.0-5.0 keV) and C-Band (5.0-12.1 keV). The hardness ratio was computed as the ratio of the addition of B-Band and C-Band to A-Band (Figure \ref{fig:figure2}). The observations in the red box in Figure \ref{fig:figure2} are the observations selected for further analysis. These observations were selected as they have the highest reflection strength and position in the HID. All selected observations are listed in Table~\ref{tab:table2}.

\begin{table*}
 		\caption{Summary of the observations analyzed in the present work.}
	\begin{tabular}{l c c c c c}
	    \hline
		\hline
		Spectrum No. & MJD & Observation ID & Observation Date & Exposure & Count Rate\\
  & & & (DD-MM-YYYY) & (s) & (cts/s)\\\hline
  Spectrum 1 & 51464 & 40124-01-06/07-00 & 13-10-1999 & 2384.0 & $1094 \pm 1.1$ \\ 
  Spectrum 2 & 51465 & 40124-01-08/09/10-00 & 14-10-1999 & 7168.0 & $1252 \pm 1.0$\\
  Spectrum 3 & 51466 & 40124-01-11-00 & 15-10-1999 & 1376.0 & $1447 \pm 2.0$\\
  Spectrum 4 & 51467 & 40124-01-12/13-00 & 16-10-1999 & 2956.0 & $2552 \pm 2.5$ \\
  Spectrum 5 & 51468 & 40124-01-14-00 & 17-10-1999 & 1408.0 & $2037 \pm 2.8$ \\
  Spectrum 6 & 51469 & 40124-01-16/17-00, 40122-01-01-00/01/02/03 & 18-10-1999 & 16820.0 & $2214 \pm 1.3$ \\
  Spectrum 7 & 51471 & 40124-01-18/19/20-00 & 20-10-1999 & 6624.0 & $1257 \pm 1.0$ \\
  Spectrum 8 & 51472 & 40124-01-21-00 & 21-10-1999 & 400.0 & $1284 \pm 2.4$  \\
  Spectrum 9 & 51473 & 40124-01-15/22/23-00, 40124-01-15/23-01 & 22-10-1999 & 4864.0 & $1294 \pm 1.0$ \\ 
  Spectrum 10 & 51474 & 40124-01-24/25-00, 40124-01-15-02/03 & 23-10-1999 & 4528.0 & $1743 \pm 1.4$ \\ 
  Spectrum 11 & 51475 & 40124-01-26/27-00 & 24-10-1999 & 8464.0 & $1453 \pm 1.2$ \\ 
  Spectrum 12 & 51476 & 40124-01-28-00/01 & 25-10-1999 & 4016.0 & $1406 \pm 1.5$ \\ 
  Spectrum 13 & 51477 & 40124-01-29-00 & 26-10-1999 & 3600.0 & $1392 \pm 1.8$ \\ 
  Spectrum 14 & 51478 & 40124-01-30/31-00, 40122-01-02-00 & 27-10-1999 & 9616.0 & $1369 \pm 0.7$ \\ 
  Spectrum 15 & 51479 & 40124-01-32-00 & 28-10-1999 & 1712.0 & $1563 \pm 2.4$ \\ 
  Spectrum 16 & 51480 & 40124-01-33-00/01 & 29-10-1999 & 880.0 & $2320 \pm 1.7$ \\ 
  Spectrum 17 & 51481 & 40124-01-34-00/01/02, 40124-01-35-01/02 & 30-10-1999 & 14930.0 & $1436 \pm 1.0$ \\ 
  Spectrum 18 & 51482 & 40124-01-35-00, 40124-01-36-01 & 31-10-1999 & 10940.0 & $1364 \pm 1.0$ \\ 
  Spectrum 19 & 51483 & 40124-01-36/37-00 & 01-11-1999 & 10540.0 & $1408 \pm 1.0$ \\ 
  Spectrum 20 & 51484 & 40124-01-37-00/01, 40124-01-38-00 & 02-11-1999 & 14430.0 & $1285 \pm 0.8$ \\ 
  Spectrum 21 & 51485 & 40124-01-38-01, 40124-01-39-00 & 03-11-1999 & 8624.0 & $958 \pm 0.9$ \\ 
  Spectrum 22 & 51486 & 40124-01-40-00/01 & 04-11-1999 & 2528.0 & $839 \pm 1.2$ \\ 
  Spectrum 23 & 51487 & 40124-01-41-00 & 05-11-1999 & 3008.0 & $814 \pm 1.6$ \\ \hline
		\hline
	\end{tabular}
	\label{tab:table2}
 \end{table*}

\begin{figure*}
\caption{Evolution of different physical parameters throughout all the selected observations. The plotted uncertainties correspond to the 90\% confidence level for one relevant parameter ($\Delta\chi^2 = 2.71$). The black and red data points and error bars represent the parameter values obtained from fitting {\tt relxill} and {\tt relxillD} models respectively. The green and blue stars represent the observations that were analysed in the sample sets G1 and G2, respectively.}
    \includegraphics[width=3.3in,angle=0,trim=0 34 0 0,clip]{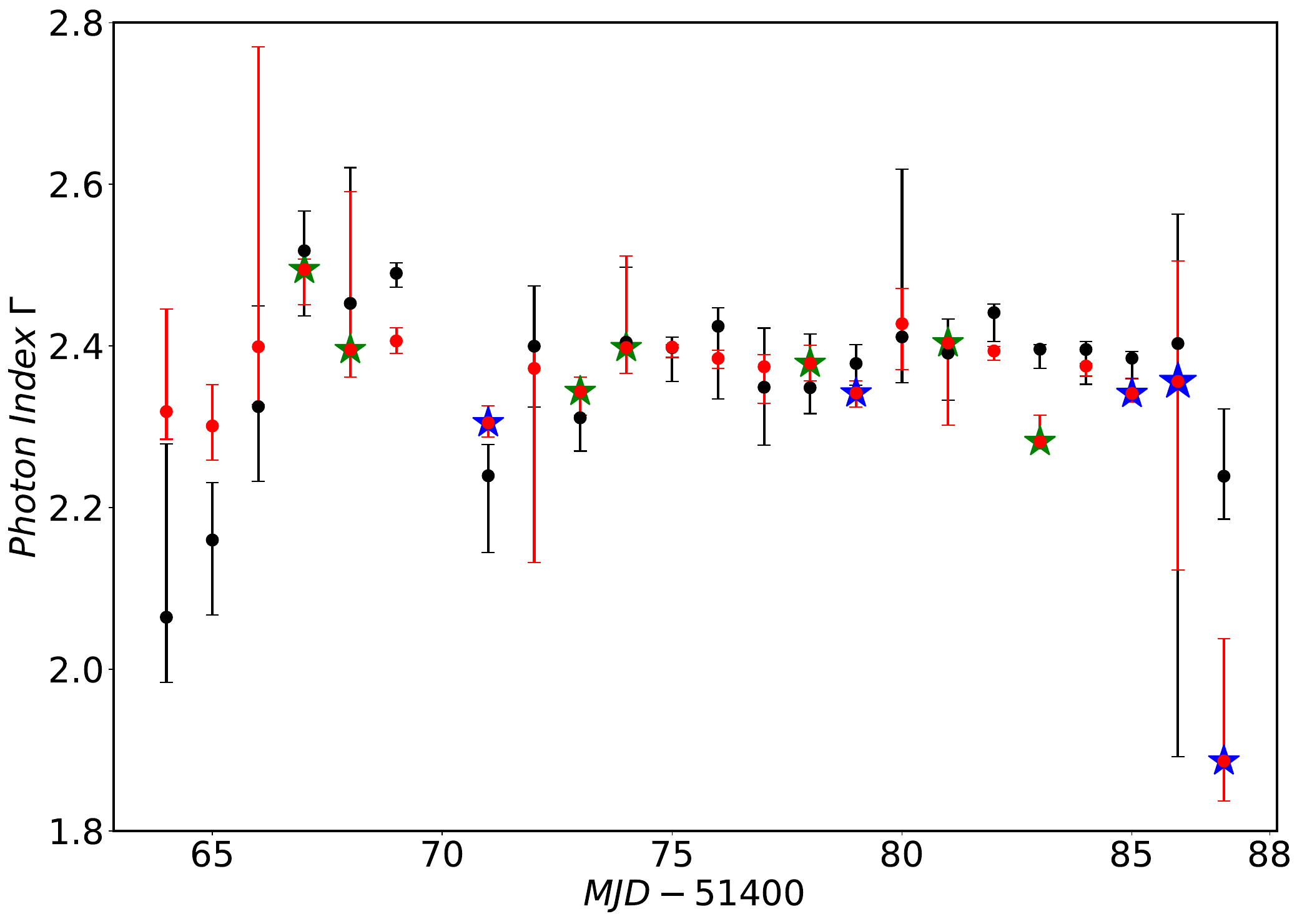}
        \includegraphics[width=3.3in,angle=0,trim=0 34 0 0,clip]{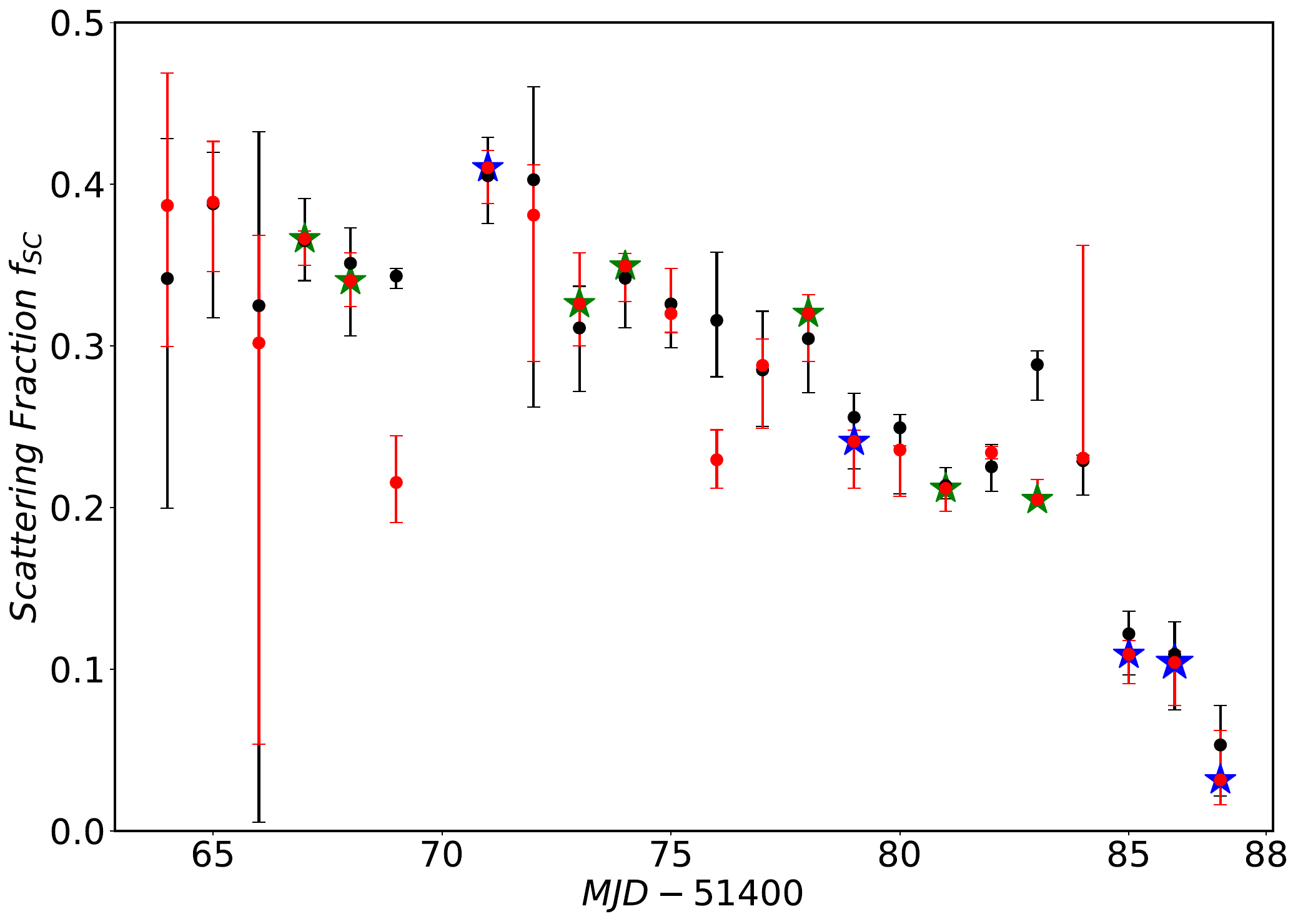}
                \includegraphics[width=3.3in,angle=0,trim=0 0 0 0,clip]{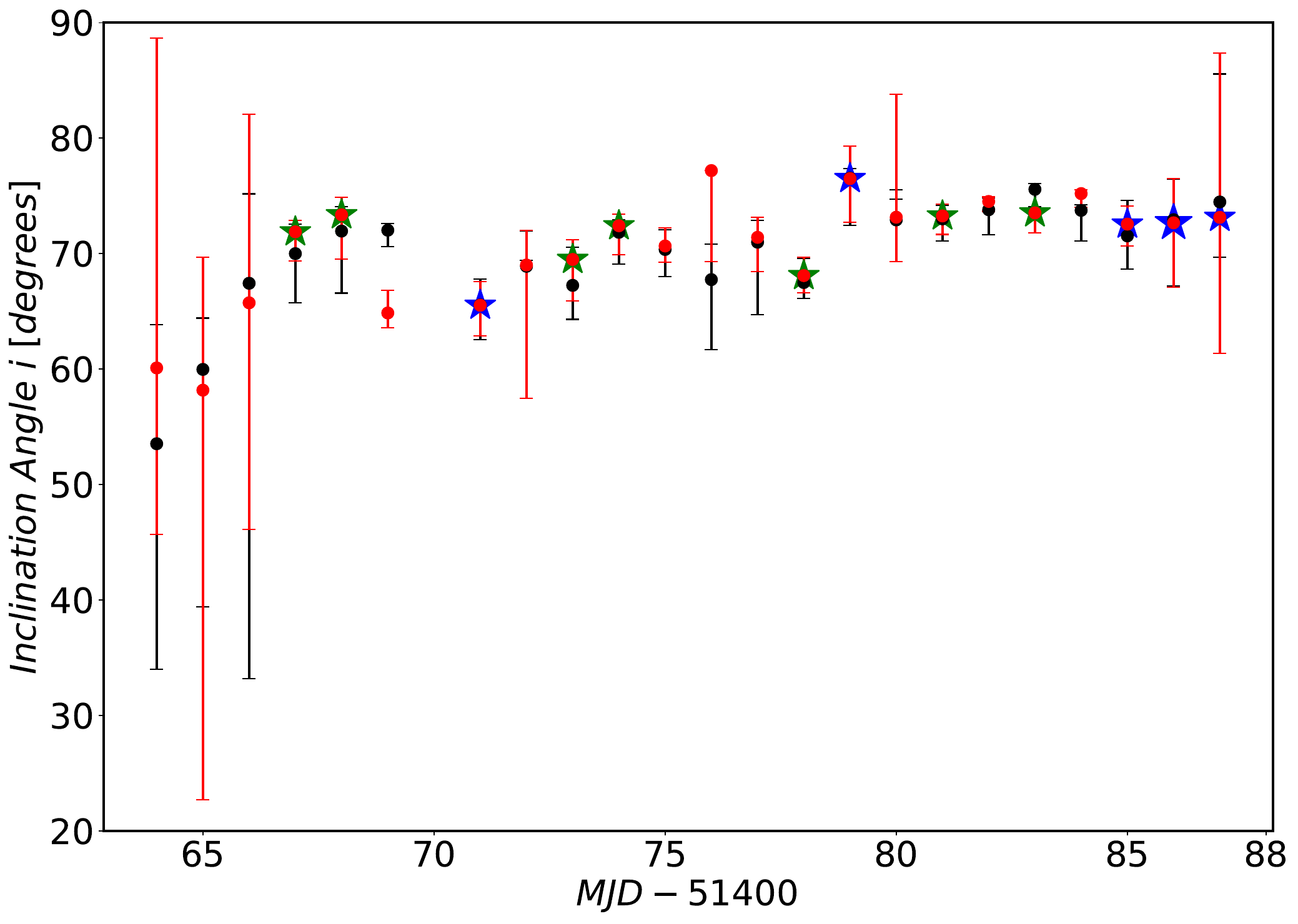}
    \includegraphics[width=3.3in,angle=0,trim=0 0 0 0,clip]{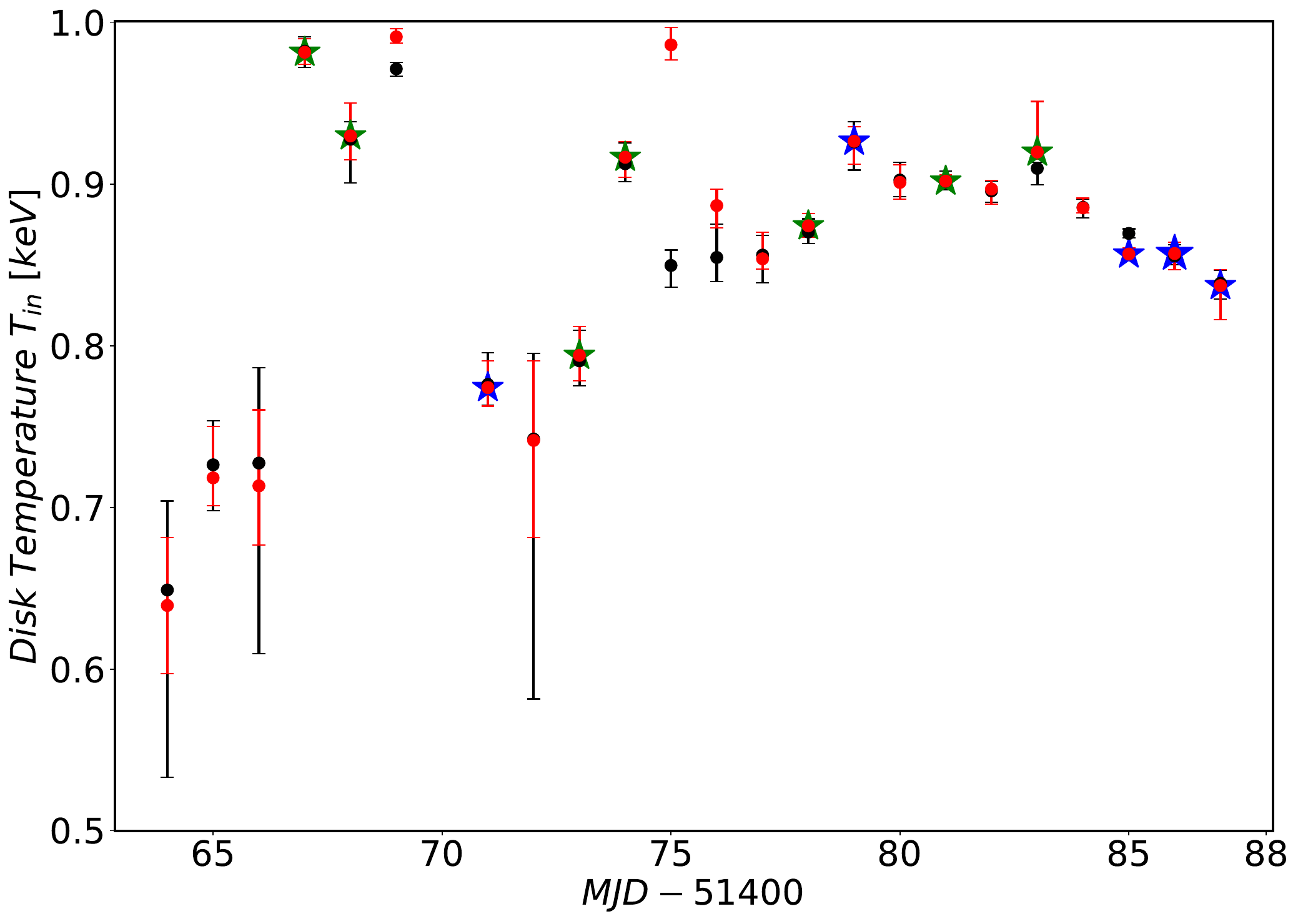}
  \label{fig:figure3}
\end{figure*}

\begin{figure*}
\caption{Evolution of different physical parameters throughout all the selected observations. The plotted uncertainties correspond to the 90\% confidence level for one relevant parameter ($\Delta\chi^2 = 2.71$). The black and red data points and error bars represent the parameter values obtained from fitting {\tt relxill} and {\tt relxillD} models respectively. The orange cross represents those parameter values that remained unconstrained. The green and blue stars represent the observations that were analysed in the sample sets G1 and G2, respectively.}
    \includegraphics[width=3.3in,angle=0,trim=0 34 0 0,clip]{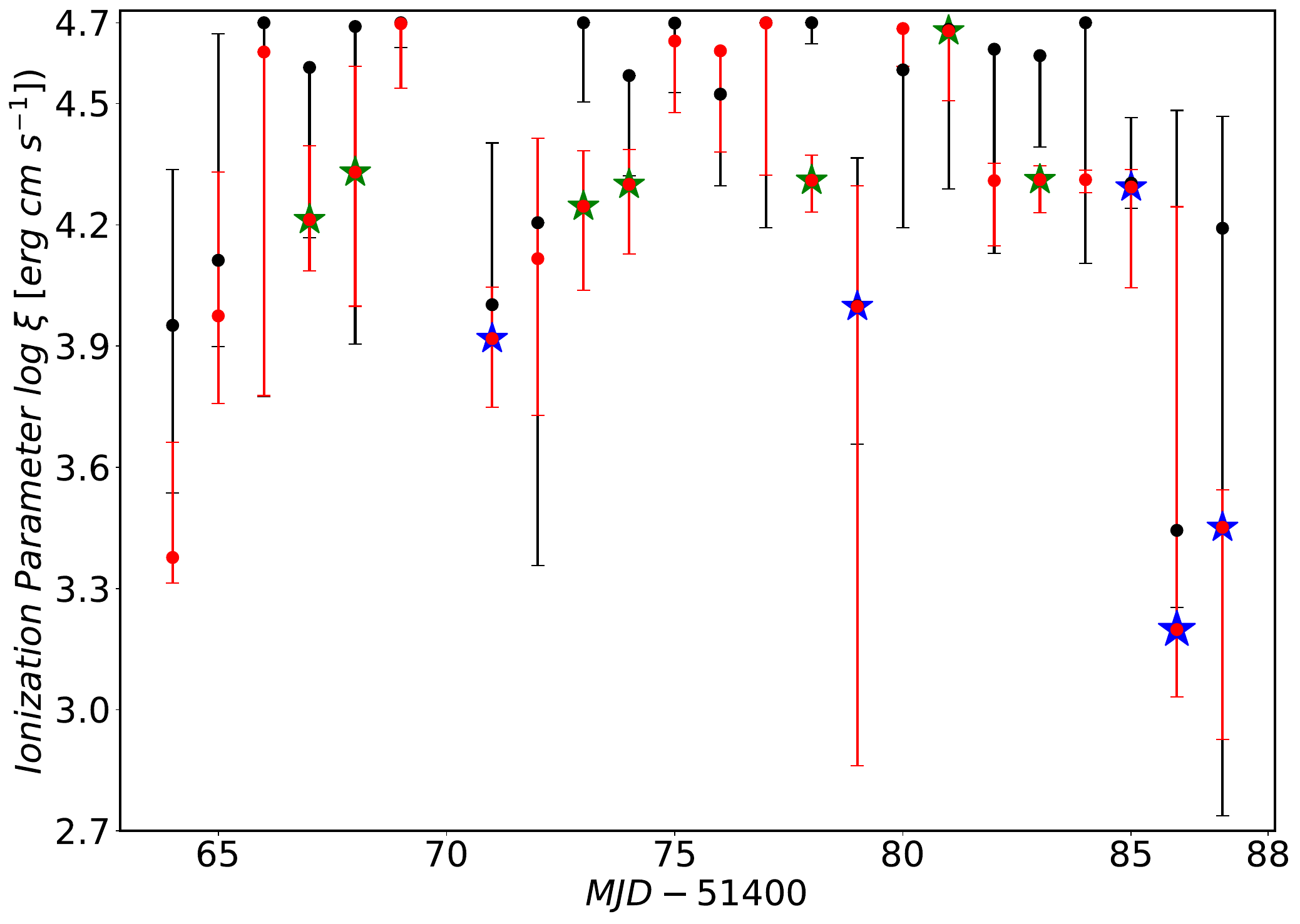}
                        \includegraphics[width=3.3in,angle=0,trim=0 34 0 0,clip]{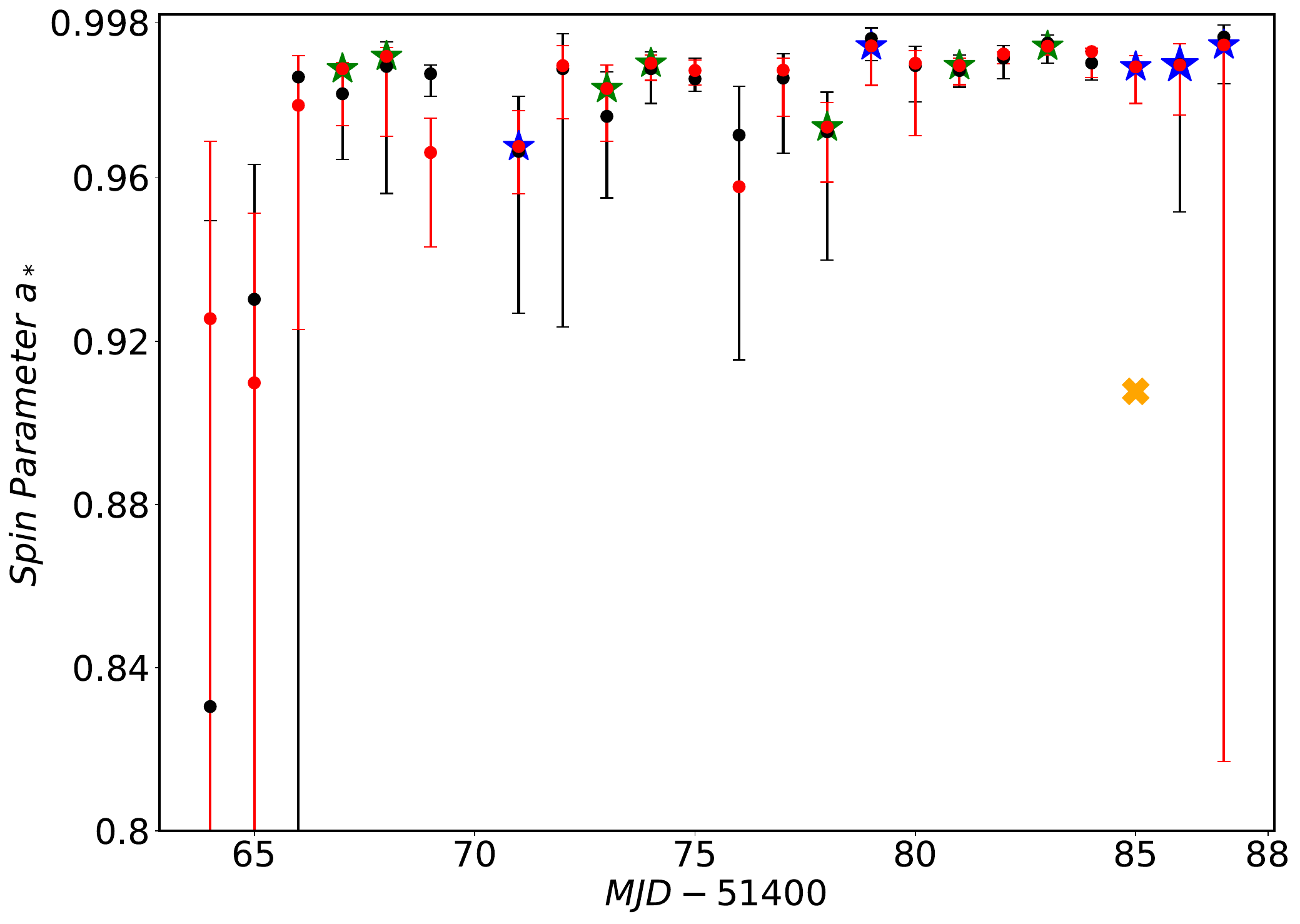}
                                \includegraphics[width=3.3in,angle=0,trim=0 0 0 0,clip]{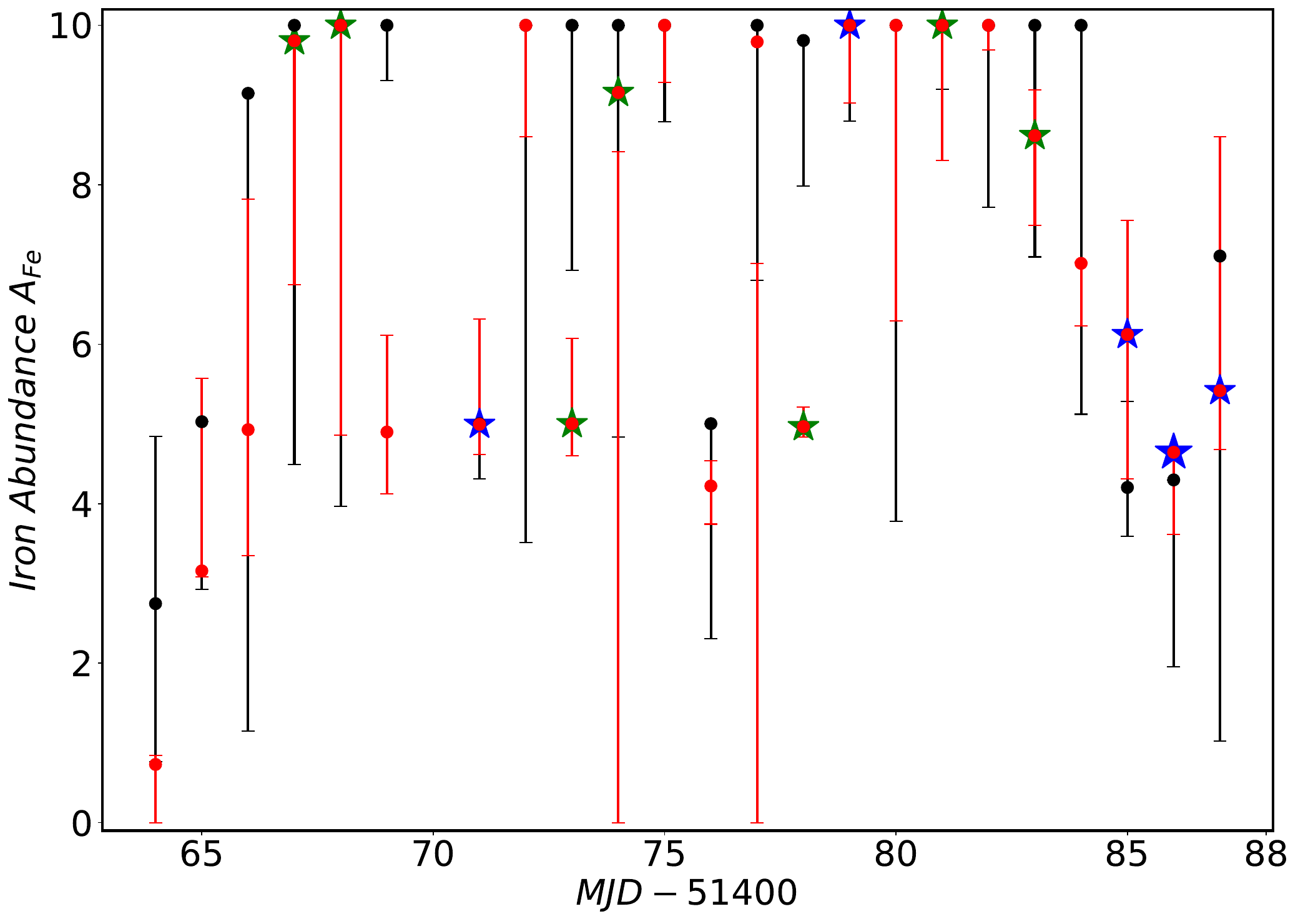}
    \includegraphics[width=3.3in,angle=0,trim=0 0 0 0,clip]{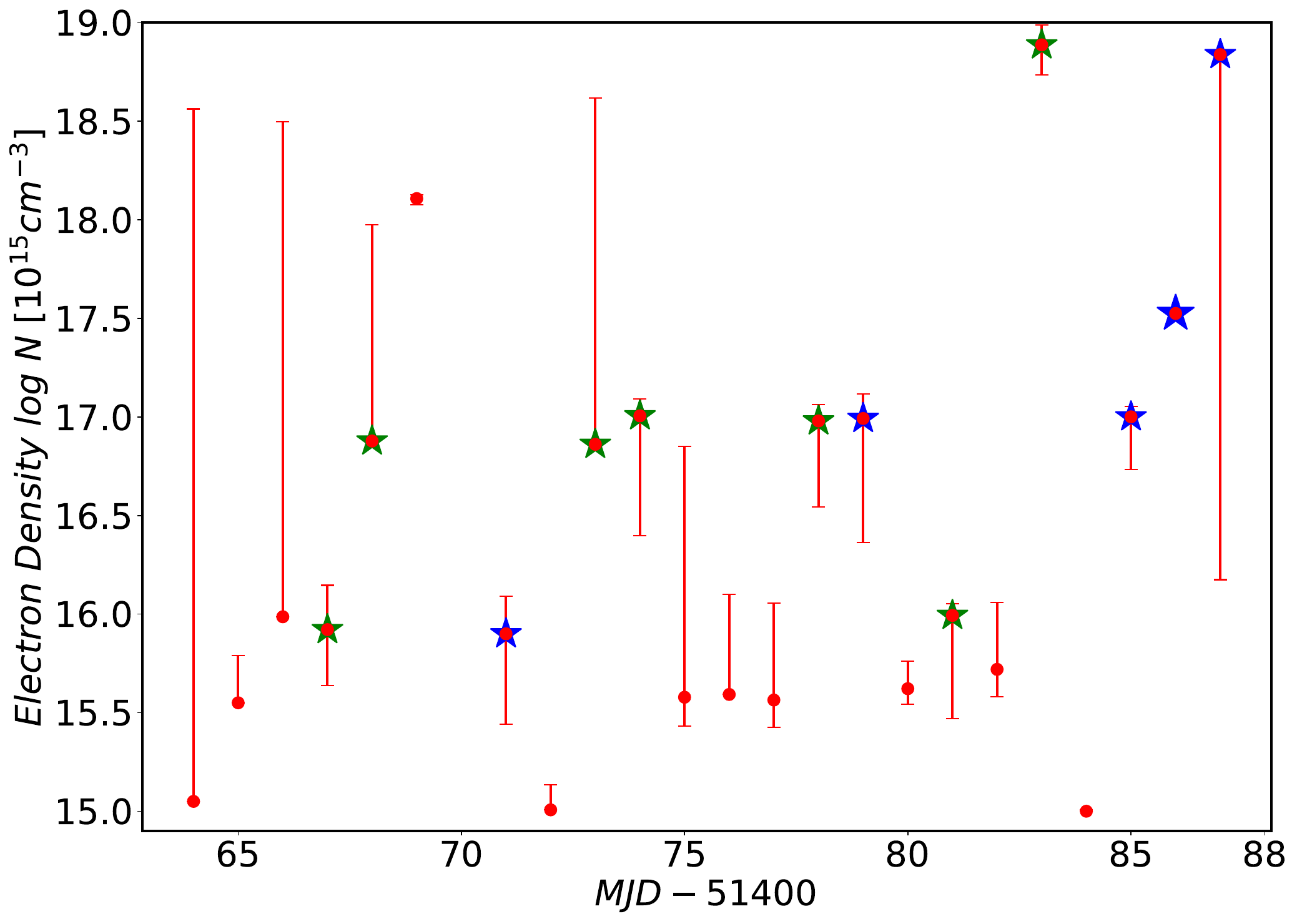}
   \label{fig:figure4}
\end{figure*}

\begin{figure}
\caption{Goodness-of-fit in terms of reduced $\chi^2$ throughout all the selected observations. The black and red data points are the $\chi^2$ values obtained from fitting the {\tt relxill} model with 66 degrees of freedom and {\tt relxillD} model with 65 degrees of freedom, respectively. The green and blue stars represent the observations that were analysed in the sample sets G1 and G2, respectively.}
    \includegraphics[width=0.45\textwidth]{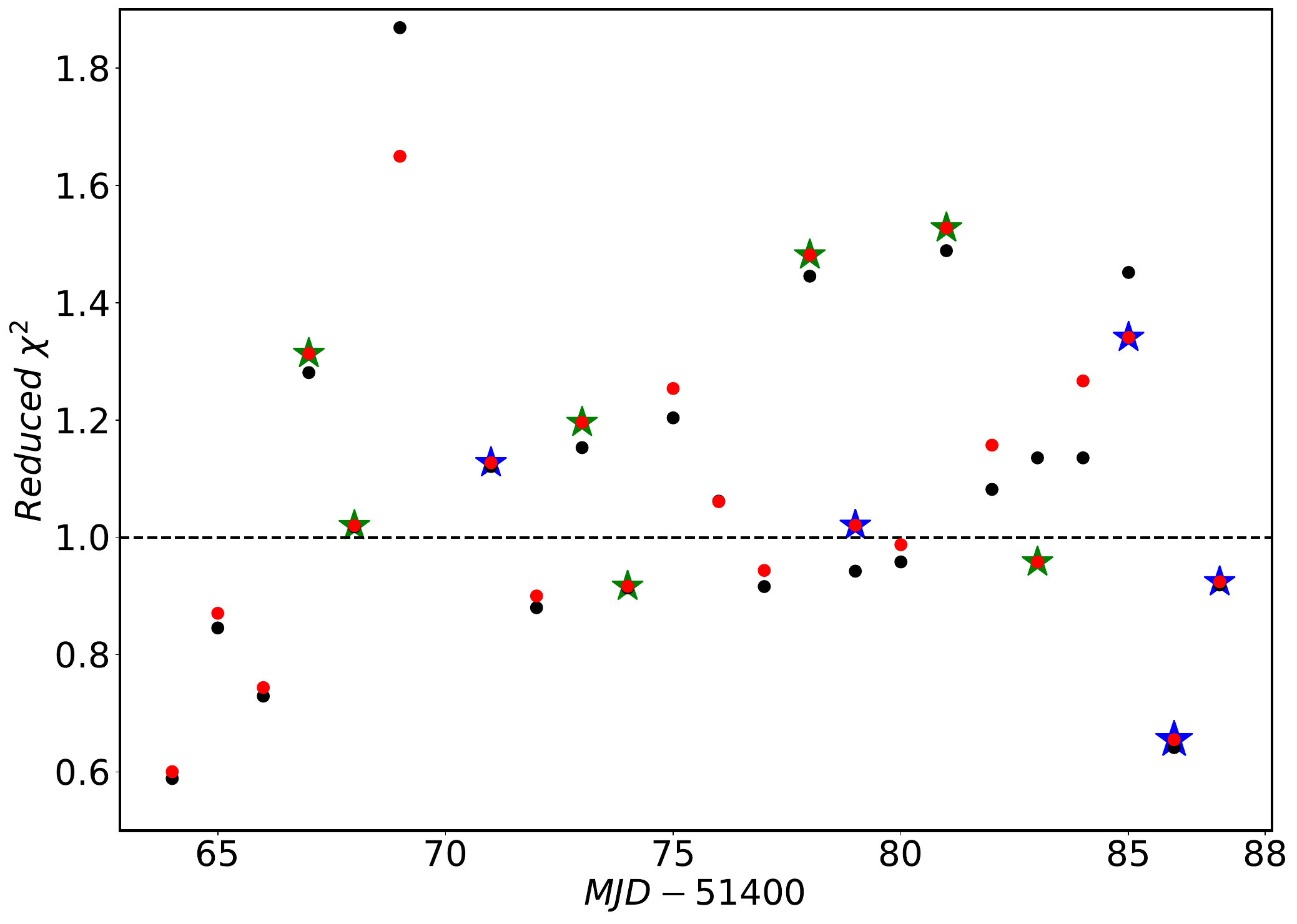}
   \label{fig:figure5}
\end{figure}

We continue the analysis of the selected observations.
A self-consistent model that accounts for the impact of coronal Comptonization on the reflection spectrum and calculates the continuum from the corona directly from the thermal spectrum is {\tt simplcut} \citep{2017ApJ...836..119S}. This model has four parameters: the photon index $\Gamma$, the scattering fraction ${f}_{\mathrm{SC}}$, the high-energy cutoff $E_{\rm cut}$, and the reflection fraction ${R}_{\rm{F}}$. The {\tt simplcut} kernel employs a redistribution function to model the Comptonization of a general population of seed photons entering the corona (e.g. \citealp{2015MNRAS.448..703W}) The scattering is not generically symmetric but includes both up- and down-scattering, for the fraction of photons that are scattered. In XSPEC notation, the total model can be written as {\tt tbabs}*{\tt simplcut}*({\tt diskbb} + {\tt relxill}). We assumed that the emissivity profile of the reflection spectrum is described by a simple power law and therefore we have to fit only one parameter, the emissivity index $q$. The reflection fraction $R_{\rm f}$ was frozen to $-1$, so the output of {\tt relxill} is only a reflection spectrum and {\tt simplcut}*{\tt diskbb} describes both the thermal spectrum from the disk and the Comptonized spectrum from the corona. The spin parameter $a_*$ was left free in the fit and the inner edge of the accretion disk was set to the ISCO radius. The iron abundance $A_{\rm Fe}$, the inclination angle of the disk with respect to the line of sight $i$, and the ionization parameter $\xi$ were all free in the fit. This model provides a good fit for all observations. 
We also considered the scenario with higher electron density, allowing us to directly fit the density of the accretion disk in logarithmic terms log$N$, and we replaced {\tt relxill} with {\tt relxillD}. 
The model then read {\tt tbabs}*{\tt simplcut}*({\tt diskbb} + {\tt relxillD}). 
On the whole, we found consistent measurements of the parameters between the fits obtained with {\tt relxill} and {\tt relxillD}. All the parameters, as they vary throughout the observations, are plotted in Figure \ref{fig:figure3} and Figure \ref{fig:figure4}. Figure \ref{fig:figure5} shows the reduced $\chi^2$ of these fits. Figure \ref{fig:figure5a} is a representative spectral fit of one spectrum of the total set to illustrate the components.

 \begin{figure}
\caption{Histogram of the reduced $\chi^2$ throughout all the selected observations. The black and red bars represent the number of observations in different bins of the reduced $\chi^2$ values obtained from fitting the {\tt relxill} model with 66 degrees of freedom and {\tt relxillD} model with 65 degrees of freedom, respectively. The blue line represents the theoretical chi-square distribution curve with 66 degrees of freedom.}
  \includegraphics[width=0.48\textwidth]{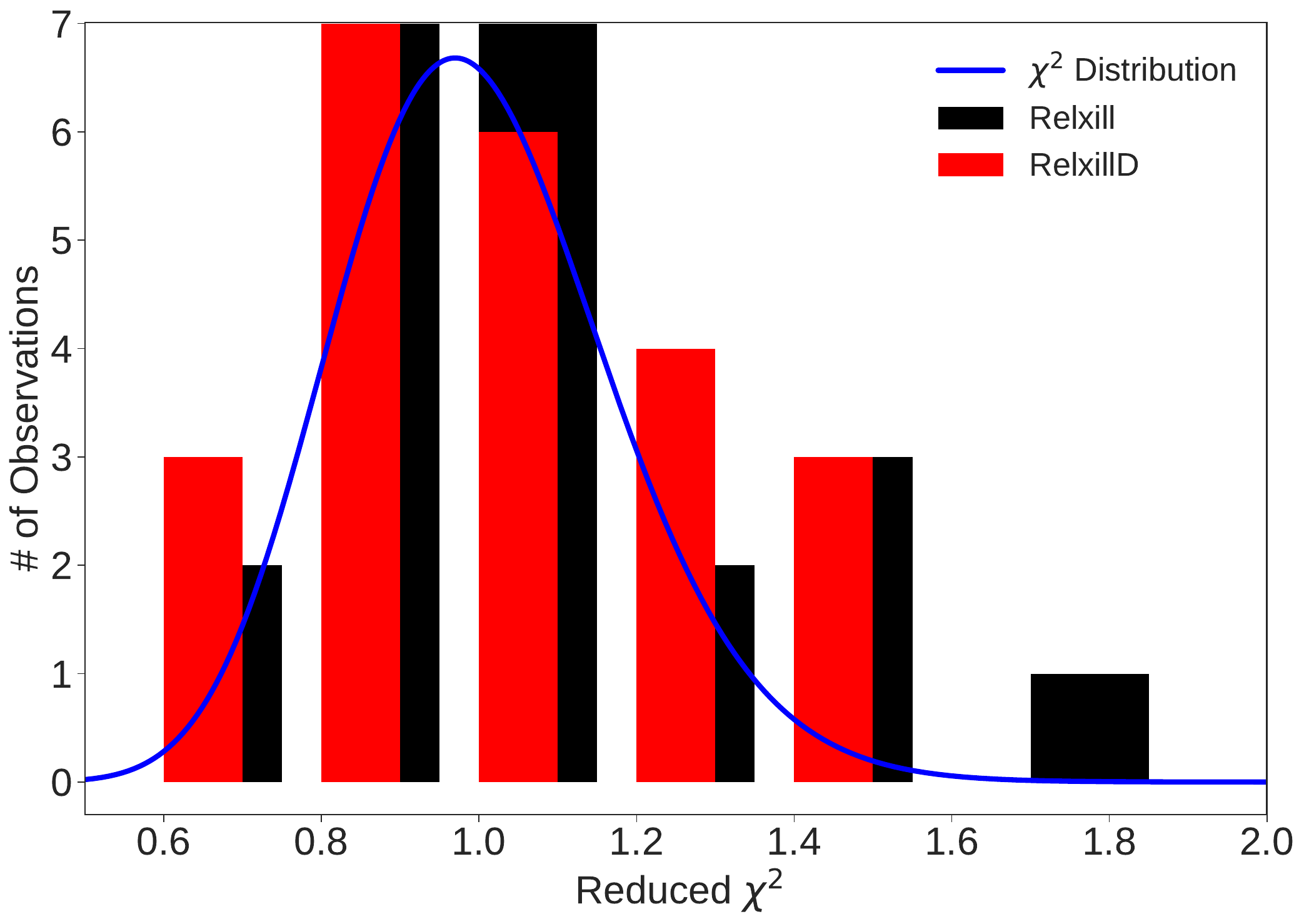}
   \label{fig:figure5b}
\end{figure}

    \begin{figure}
\caption{The top panel shows Spectrum 20 as fitted with the {\tt relxill} model. Here the black line is the total spectrum, the green line represents a sum of the scattered and transmitted disk component, and the purple line represents the scattered and transmitted reflection component. The middle and bottom panels show the $\Delta\chi$ when the {\tt relxill} and {\tt relxillD} model formulations were employed to model the spectrum, respectively. }
    \includegraphics[width=0.45\textwidth]{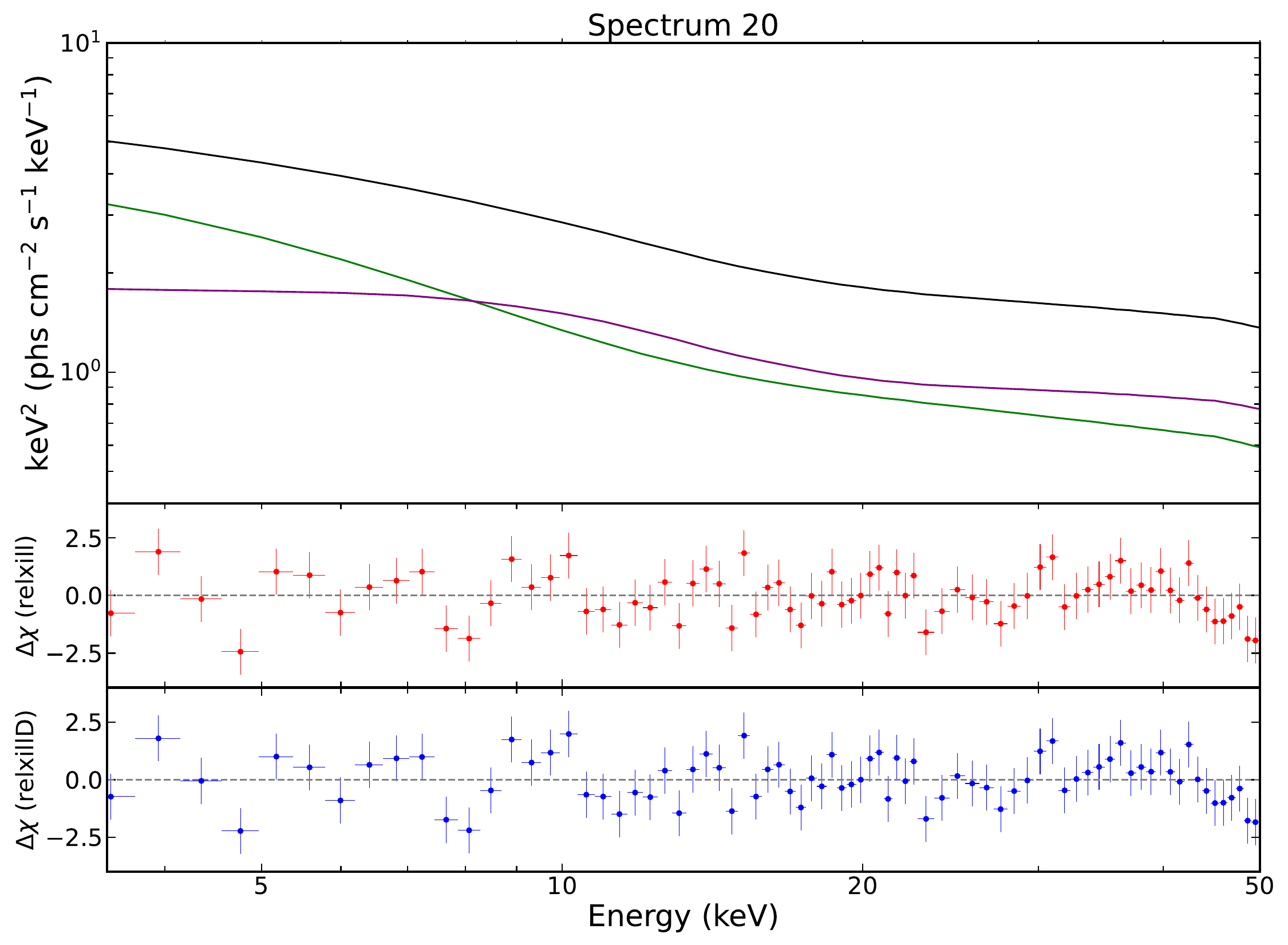}
   \label{fig:figure5a}
\end{figure}

 \begin{figure}
\caption{The top panel shows Spectrum 20 as fitted with the model formulation: {\tt tbabs}*(({\tt simplcut}*{\tt diskbb}) + {\tt relxill}). Here the black line is the total spectrum, the green line represents the sum of the scattered and transmitted disk component, and the purple line represents the reflection component. The middle and bottom panels show the $\Delta\chi$ when the {\tt relxill} and {\tt relxillD} model formulations were employed to model the spectrum, respectively.}
    \includegraphics[width=0.45\textwidth]{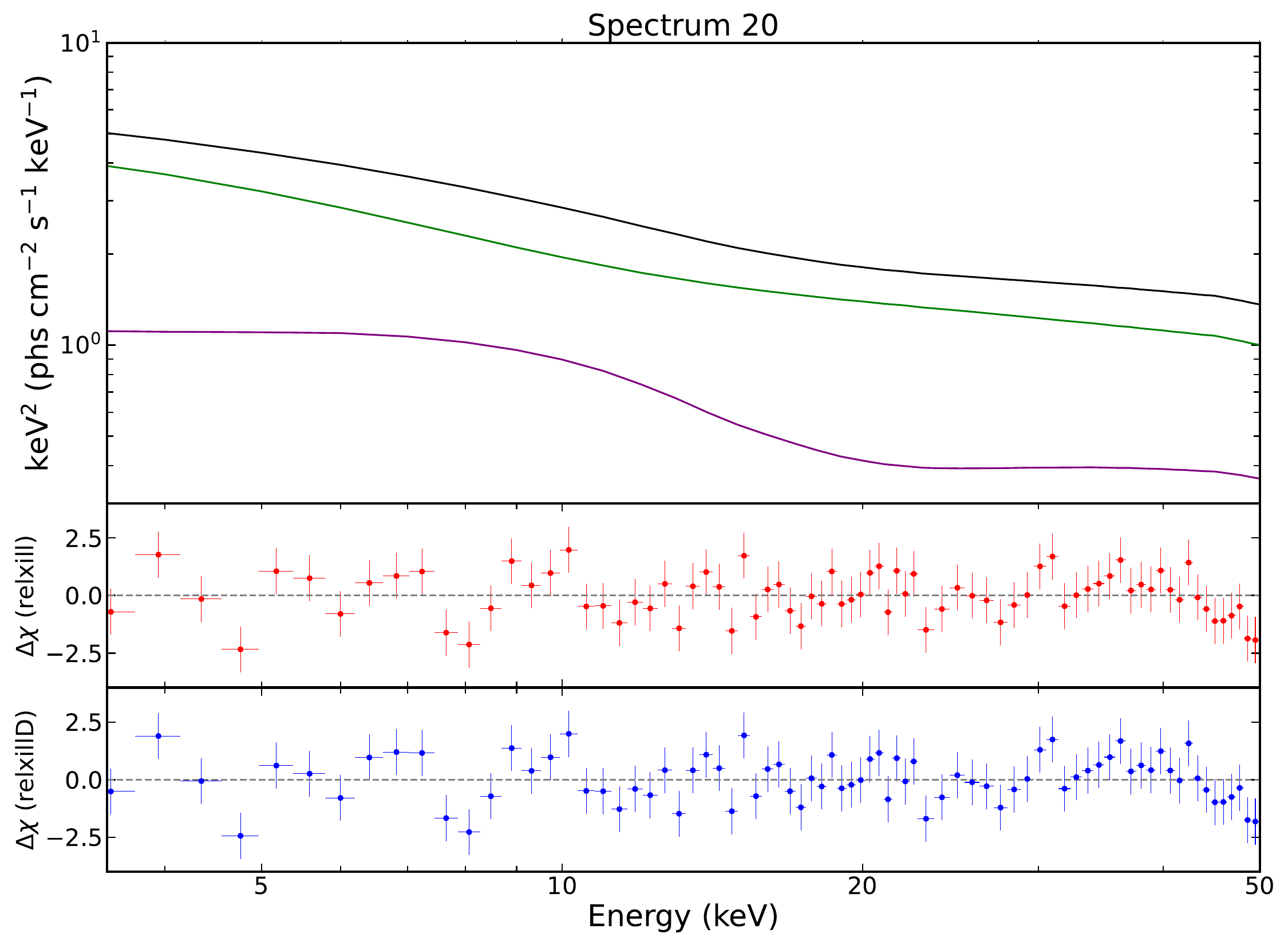}
   \label{fig:figure8}
\end{figure}

After fitting spectral models to the data and treating all data sets independently, we ultimately produced a simultaneous fit linking several of the invariant parameters for a selected set from 23 observations. This small set of spectra (G1 Set) was selected taking into account various factors. The photon count rate of the 23 spectra was taken into account and only the spectra with more than 1000 counts/s were chosen. The spectrum whose hardness was above 1.3 and the reflection strength was above 16 were selected and thus a few more spectra were removed from the shortlisted set. Furthermore, we considered only the spectra in which the best-fit returned constrained parameters in the independent fit. The G1 Set eventually included 7 spectra to be used in the simultaneous fitting. Figure \ref{fig:figure6a} shows the data to model ratios for this set of spectra. Using the models specified above, in this simultaneous fit, the spin parameter $a_*$, the inclination angle $i$, and the iron abundance A$_{\rm Fe}$ are tied among all spectra, whereas the other parameters are allowed to vary among the observations. The best-fit parameter values for both model formulations, {\tt relxill} and {\tt relxillD} are listed in Tables~\ref{table:table3}~and~\ref{table:table4}, respectively, with the errors corresponding to the 90\% confidence level. The unfolded spectra with residuals, for both the model formulations aforementioned, are shown in Figures~\ref{fig:figure6}. We noted that all the spectrum in G1 Set had high values of the ionization parameter. To take into account the spectrum that might have a more prominent Fe-K$\alpha$ feature, we performed another simultaneous fit on another set of spectrum (G2 Set). For the G2 Set, we selected only those spectra, from the 23, whose {\tt relxill} or {\tt relxillD} fits returned a value of ionization parameter {log $\xi$} less than 4.3. We discarded the spectra whose exposure was less than 500 seconds and the ones whose individual spectral fitting did not return constrained parameters. We performed the simultaneous fitting as described above. The best-fit parameter values for both model formulations, {\tt relxill} and {\tt relxillD} are listed in Tables~\ref{table:table5}~and~\ref{table:table6}, respectively, with the errors corresponding to the 90\% confidence level. The unfolded spectra with residuals, for both the model formulations aforementioned, are shown in Figures~\ref{fig:figure7}.  We also explored a slightly different model formulation to exclude the impact of coronal comptonization on the reflection spectrum. In XSPEC notation, the model then was {\tt tbabs}*(({\tt simplcut}*{\tt diskbb}) + {\tt relxill}/{\tt relxillD}). Only the scattering fraction ${f}_{\mathrm{SC}}$ and the inner disk radius obtained from {\tt diskbb} were inconsistent with the previous fits. Both these parameter values increased by 40\%. However, every other parameter value remained consistent with the previous fits and the reduced $\chi^2$ values also did not change significantly. Figure \ref{fig:figure8} is a representative spectral fit of one spectrum of the total set to illustrate the components. The results are discussed in the next section.


\section{Discussion and conclusions}
\label{sec:4}

We carried out broadband spectral analysis of 23 distinct observations of XTE~J1859+226 as observed by \textsl{RXTE} to determine the spin parameter of the compact object of this source via XRS. To obtain this result we fit the reflection features of the spectra with {\tt relxill}/{\tt relxillD}, {\tt diskbb}, and {\tt simplcut}. We note that both the reflection 
models provided a good fit for all the spectra. 
The best-fit parameter values obtained from the independent fits are plotted in Figures \ref{fig:figure3} and \ref{fig:figure4} at a 90\% CL (statistical). We note that the obtained values of spin parameter $a_*$ are close to the maximum possible value of the parameter. The values of the spin parameter plotted in Figure \ref{fig:figure4} seem to be systematically increasing with time. This is a topic of further investigation whether this could be an indication of an evolving disk geometry through the observations. However, this is beyond the scope of this work. The spin parameter $a_*$ is $0.986^{+0.001}_{-0.004}$ and $0.987 \pm 0.003$, as inferred from the simultaneous fitting of the G1 Set of spectra, for {\tt relxill} and {\tt relxillD} models, respectively. The value obtained for the inclination angle of the accretion disk, $i$, is $71.2_{-0.8}^{+0.4}$ degrees and $71.7^{+0.9}_{-0.3}$ degrees, respectively. The iron abundance A$_{\rm Fe}$ obtained is $10.0_{-0.4}$ and $7.12^{+0.28}_{-0.18}$, respectively. For the G2 Set, the values of $a_*$ from two models are $0.981_{-0.007}^{+0.006}$ and $0.982_{-0.007}^{+0.006}$; $i$ is $68.4_{-1.4}^{+1.9}$ degrees and $69.1_{-0.8}^{+0.5}$ degrees; $A_{\rm Fe}$ is $10.0_{-1.7}$ and $5.23_{-0.25}^{+0.5}$.
This measurement of spin is consistent with the spin distribution in X-ray binaries~\citep{2022ApJ...929L..26F}. The high value of the inclination angle of the disk is found consistent with suggested values of the same in previous studies~\citep{2011MNRAS.413L..15C,2022MNRAS.517.1476Y}. The reduced $\chi^2$ is $<$ 1 for some of the individual fits as reported in the aforementioned tables. It is possible that the systematic uncertainties included have led to this overfitting. However, this is important for instrumental and calibration uncertainty and thus can not be reduced. Figure~\ref{fig:figure5b} presents the histogram of the reduced $\chi^2$ values obtained throughout all the selected observations. For comparison, we have plotted the theoretical chi-square distribution curve for 66 degrees of freedom and it is consistent with the histogram.

Our reported value of $a_*$ is clearly inconsistent with the value reported by \citet{2022MNRAS.517.1469M} (see Section \ref{sec:Intro}). It is worth mentioning that the respective authors have reported that they found one clear QPO triplet in the observation ID 40124-01-07-00 (Spectrum 1, Table \ref{tab:table2}) which is also one of the observations we have made use of in our study (Table \ref{tab:table2}). Such discrepancy between the spin measurements obtained via XRS and RPM has also been noted in the case of another LMXB source, GRO~J1655-40. The spin measurements via RPM rely strongly on two parameters: the mass of the BH and the radial coordinate of the oscillation. A stringent constraint on these parameters can be obtained using a simultaneous observation of a low-frequency QPO and twin high-frequency QPOs, with the assumption that these QPOs arise at a given radius. \citet{2016ApJ...825...13S} demonstrated for the case of GRO~J1655-40 that if this assumption that the twin high-frequency QPOs and the simultaneously observed low-frequency QPO being observed at a common radius is abandoned, then the $a_*$ prediction made by this model is in good agreement with the values that are obtained via XRS.

In our fits, the inner edge of the accretion disk in {\tt relxill}/{\tt relxillD} is set to the ISCO radius, and therefore it can range from 1.23~$R_{\rm g}$ ($a_* = 0.998$) to 9~$R_{\rm g}$ ($a_* = -1$), where $R_{\rm g}=G_{\rm N}M/c^2$ is the gravitational radius of the BH. In the RPM, the oscillation can be at the ISCO or at a larger radius. If the disk is truncated at a radial coordinate larger than 9~$R_{\rm g}$, this is a problem for the XRS measurement. One of the observations considered in our final sample is the observation with the QPO triplet considered by \citet{2022MNRAS.517.1469M}: for that observation, \citet{2022MNRAS.517.1469M} inferred a characteristic radius of 6.85~$R_{\rm g}$, which is in the range of our reflection model. Fig.~4 in \citet{2022MNRAS.517.1469M} reports the centroid frequencies of the Type-C QPOs of the source and they are in the range $\sim$1-10~Hz. A few QPOs have a characteristic frequency centered around 1~Hz, which would correspond to a nodal precession radius exceeding 9~$R_{\rm g}$ according to the RPM. However, none of those observations have been used in our work.

\cite {10.1111/j.1745-3933.2009.00693.x} and \cite{10.1111/j.1365-2966.2011.18860.x} presented a physical model for the QPOs where a hot, geometrically thick accretion flow is undergoing the Lense-Thirring precession and is misaligned to the BH spin axis. To account for the effects of a geometrically thin disk surrounding this hot, geometrically thick accretion flow, \cite{Bollimpalli_2022} performed general relativistic magnetohydrodynamics simulations. They concluded that the outer thin disk affects the precession of the inner thick disk by significantly reducing its rate, which, in turn, has direct implications for the truncation radius estimation and the observed QPO frequency range. 
Another possible way to obtain a truncated disk is by fitting the spectra with two Comptonisation and reflection components, mimicking the effect of a stratified extended corona. The extra curvature this creates in the spectrum has a similar effect to the red wing of a very broad iron line and this would impact the estimated spin parameter (e.g. \citealt{Frontera_2001, Shidatsu_2011, Zdziarski_2021}).

Indeed these timing-based estimates are close to the spin distribution emerging from GW data analysis. On the other hand, in the case of MAXI~J1820+070, the various values obtained in the literature are not fully consistent with each other. However, \citet{2021MNRAS.508.3104B} does mention that the RPM tends to introduce a bias. This happens due to the use of broad features instead of QPO peaks and it underestimates the value of the spin parameter, thus their reported value should be regarded as a lower limit. The reliability of the RPM model depends on the interpretation of QPOs. The measured BH spins, assuming the QPOs are the dynamic frequencies, are consistent with values obtained from reflection spectroscopy~\citep{2020ApJ...889L..36M,2021ApJ...909...63L,2023arXiv230518249Z}.

\citet{2023ApJ...946...19D} discuss in detail how the overall distribution of spin measurements conducted via XRS is stacked near high values. All the available reflection models present a number of simplifications that can, at different levels, have an impact on the precision measurements of BHs. It is important to remember for both these methods of measurement that the discrepancy probably boils down to the assumptions inherent in the two models and these models can give an estimate of the spin assuming only that framework.

\section*{Acknowledgements}
This work was supported by the National Natural Science Foundation of China (NSFC), Grant No.~12250610185, 11973019, and 12261131497, the Shanghai Municipal Education Commission, Grant No.~2019-01-07-00-07-E00035, the Natural Science Foundation of Shanghai, Grant No.~22ZR1403400, and Fudan University, Grant No. JIH1512604.
GM acknowledges also the support from the China Scholarship Council (CSC), Grant No. 2020GXZ016647.

\section*{DATA AVAILABILITY}
This work used the data and software provided by the High Energy Astrophysics Science Archive Research Center (HEASARC), which is a service of the Astrophysics Science Division at NASA/GSFC.



\bibliographystyle{mnras}
\bibliography{Cite.bib} 

\appendix
\section{Estimated Parameter Values}

As mentioned in Section \ref{sec:3}, we selected 7 out of the 23 spectra, and named it the G1 Set, to perform the simultaneous fitting. Another 5 spectra were selected from the 23 in the G2 set with a different set of criteria for the same. The parameter values obtained when these spectra were independently fit are noted in the top half of the tables in this section. The bottom half has the values of the parameters when these spectra were fit together as explained in Section \ref{sec:3}. Tables \ref{table:table3} and \ref{table:table5} has the values obtained from using the model {\tt relxill}. Tables \ref{table:table4} and \ref{table:table6} have the values obtained from employing the model {\tt relxillD}. We have also displayed the unfolded spectra with the respective residuals for both models and both sets below.

\begin{table*}
\caption{Summary of the best-fit values from the analysis of the G1 Set of spectra of XTE J1859+226 when the emissivity profile of the disk is described by a power law. $^*$ indicates that the parameter is frozen in the fit. The reported uncertainties correspond to the 90\% confidence level for one relevant parameter ($\Delta\chi^2 = 2.71$).}
\renewcommand{\arraystretch}{2}
\setlength{\tabcolsep}{3pt}
\label{table:table3}
\rotatebox{90}{
\begin{tabular}{l|c| c| c| c| c| c | c| c| c | c| c| c }
  \hline
  \hline
 Spec & $\Gamma$ & ${f}_{\mathrm{SC}}$ & $E_{\rm cut}$ & $T_{\rm in}$ & norm & $q$ & $a_*$ & $i$ & $A_{\rm Fe}$ & log$\xi$ & norm & $\chi^2$/$\nu$ \\ 
 & & & [keV] & [keV] & & & & [deg] & & [erg~cm~$s^{-1}$] & & \\\hline
 Spec 4 & $2.52_{-0.08}^{+0.04}$ & $0.365_{-0.025}^{+0.026}$ & $236_{-102}$ & $0.982_{-0.01}^{+0.009}$ & $2194_{-153}^{+177}$ & $10_{-1.5}$ & $0.981_{-0.016}^{+0.007}$ & $70_{-4}^{+3}$ & $10_{-5}$ & $4.5_{-0.4}$ & $0.111_{-0.022}^{+0.03}$ & 84.53/66\\
    \hline
Spec 5 & $2.45_{-0.06}^{+0.16}$ & $0.351_{-0.04}^{+0.022}$ & $146^{+203}_{-44}$ & $0.928_{-0.027}^{+0.011}$ & $2074_{-223}^{+274}$ & $10_{-1.4}$ & $0.987_{-0.03}^{+0.006}$ & $71.95_{-5.4}^{+2.12}$ & $10_{-6}$ & $4.6_{-0.7}$ & $0.11_{-0.04}^{+0.05}$ & 67.18/66\\
    \hline
Spec 9 & $2.31_{-0.04}^{+0.03}$ & $0.311_{-0.03}^{+0.026}$ & $131^{+72}_{-21}$ & $0.791_{-0.016}^{+0.019}$ & $2210_{-256}^{+204}$ & $10_{-1.9}$ & $0.975_{-0.020}^{+0.011}$ & $67.3_{-2.9}^{+3}$ & $10_{-3}$  & $4.7_{-0.19}$ & $0.063_{-0.010}^{+0.022}$ & 76.09/66\\
    \hline
Spec 10 & $2.405_{-0.011}^{+0.09}$ & $0.342_{-0.03}^{+0.010}$ & $156^{+131}_{-37}$ & $0.913_{-0.011}^{+0.013}$ & $1928_{-150}^{+69}$ & $10_{-0.9}$ & $0.987_{-0.009}^{+0.004}$ & $71.85_{-2.7}^{+1.11}$ & $10_{-5}$ & $4.56_{-0.24}$ & $0.086_{-0.021}^{+0.015}$ & 60.31/66\\
    \hline
Spec 14 & $2.35_{-0.03}^{+0.06}$ & $0.305_{-0.03}^{+0.017}$ & $149^{+38}_{-22}$ & $0.870_{-0.007}^{+0.008}$ & $1720_{-147}^{+74}$ & $10_{-0.8}$ & $0.971_{-0.03}^{+0.010}$ & $67.49_{-1.4}^{+2.07}$ & $9.8_{-1.8}$ & $4.70_{-0.05}$ & $0.059_{-0.010}^{+0.016}$ & 95.38/66\\
    \hline
Spec 17& $2.39_{-0.05}^{+0.04}$ & $0.214_{-0.008}^{+0.011}$ & $201^{+155}_{-48}$ & $0.902_{-0.005}^{+0.006}$ & $2050.26_{-101.07}^{+76.3}$ & $10_{-0.9}$ & $0.986_{-0.004}^{+0.004}$ & $73.00_{-1.9}^{+1.16}$ & $10_{-0.8}$ & $4.7_{-0.3}$ & $0.066_{-0.008}^{+0.020}$ & 98.25/66\\
    \hline
Spec 19& $2.396_{-0.024}^{+0.005}$ & $0.288_{-0.022}^{+0.009}$ & $  500_{-76}$ & $0.910_{-0.010}^{+0.004}$ & $1550_{-155}^{+89}$ & $10_{-0.29}$ & $0.993_{-0.005}^{+0.002}$ & $75.6_{-1.5}^{+0.4}$ & $10_{2.9}$ & $4.619_{-0.226}$ & $0.093_{-0.005}^{+0.031}$ & 63.34/66\\  \hline
   \hline
\multicolumn{13}{c}{Joint Fitting Parameter Values}\\
   \hline
   \hline
Spec 4 & $2.502_{-0.027}^{+0.06}$ & $0.370_{-0.016}^{+0.017}$ & $196^{+128}_{-66}$ & $0.977_{-0.005}^{+0.006}$ & $2302_{-81}^{+61}$ & $9.7_{-0.5}$ &  & & & $4.35_{-0.24}$ & $0.114_{-0.007}^{+0.009}$ & \\ \cline{1-7}\cline{11-12}
Spec 5 & $2.45_{-0.04}^{+0.03}$ & $0.356_{-0.009}^{+0.024}$ & $149^{118}_{-36}$ & $0.930_{-0.008}^{+0.008}$ & $2076_{-44}^{167}$ & $10_{-0.6}$ &  & & & $4.69_{-0.24}$  & $0.103_{-0.03}^{+0.007}$ & \\ \cline{1-7}\cline{11-12}
Spec 9 & $2.194_{-0.05}^{+0.021}$ & $0.335_{-0.022}^{+0.029}$ & $74^{+9}_{-7}$ & $0.822_{-0.006}^{+0.006}$ & $2137_{-118}^{+69}$ & $10_{-0.2}$ &  \multirow{0}{*}{$0.986_{-0.004}^{+0.001}$} & \multirow{0}{*}{$71.2_{-0.8}^{+0.4}$} & \multirow{0}{*}{$10.0_{-0.4}$} & $4.31_{-0.04}^{+0.3}$ & $0.048_{-0.003}^{+0.004}$ & \multirow{0}{*}{$569.02/480$} \\\cline{1-7}\cline{11-12}
Spec 10 & $2.412_{-0.018}^{+0.016}$ & $0.344_{-0.022}^{+0.015}$ & $164^{+70}_{-35}$ & $0.912_{-0.005}^{+0.01}$ & $1926_{-70}^{+138}$ & $9.9_{-0.5}$ & & & & $4.59_{-0.12}$ & $0.085_{-0.019}^{+0.006}$ & \\ \cline{1-7}\cline{11-12}
Spec 14 & $2.325_{-0.009}^{+0.019}$ & $0.284_{-0.028}^{+0.011}$ & $128^{+26}_{-16}$ & $0.868_{-0.004}^{+0.004}$ & $1703_{-103}^{49}$ & $10_{-0.25}$ &  &  &  & $4.68_{-0.07}$ & $0.072_{-0.014}^{+0.004}$ &\\ \cline{1-7}\cline{11-12}
Spec 17& $2.386_{-0.023}^{+0.008}$ & $0.217_{-0.016}^{+0.007}$ & $187^{75}_{-28}$ & $0.900_{-0.002}^{0.003}$ & $2080_{-56}^{+34}$ & $8.699_{-0.3}^{+0.4}$ &  &  &  & $4.62_{-0.08}$ & $0.058_{-0.010}^{+0.003}$ & \\ \cline{1-7}\cline{11-12}
Spec 19& $2.395_{-0.008}^{+0.006}$ & $0.292_{-0.003}^{+0.016}$ & $500_{-104}$ & $0.916_{-0.005}^{+0.003}$ & $1510_{-17}^{31}$ & $8.19_{-0.27}^{+0.4}$ &  &  &  & $4.69_{-0.06}$ & $0.069_{-0.012}^{+0.003}$ & \\\hline\hline
\end{tabular}}
\end{table*}

\begin{table*}
\caption{Summary of the best-fit values from the analysis of the G1 Set of spectra of XTE J1859+226, when a power law describes the emissivity profile of the disk and the disk density, is left free. $^*$ indicates that the parameter is frozen in the fit. The reported uncertainties correspond to the 90\% confidence level for one relevant parameter ($\Delta\chi^2 = 2.71$).}
{\scriptsize
\renewcommand{\arraystretch}{2}
\setlength{\tabcolsep}{2.6pt}
\label{table:table4}
\rotatebox{90}{
\begin{tabular}{l| c| c| c| c| c |c| c c c |c |c |c |c}
  \hline
  \hline
   Spec & $\Gamma$ & ${f}_{\mathrm{SC}}$ & kT$_e$ & $T_{\rm in}$ & norm & $q$ & $a_*$ & $i$ & $A_{\rm Fe}$ & log$\xi$ & log N & norm & $\chi^2/\nu$ \\ 
 & & & [keV] & [keV] & & & & [deg] & & [erg cm $s^{-1}$] & & & \\\hline
Spec 4 & $2.495_{-0.04}^{+0.013}$ & $0.366_{-0.017}^{+0.005}$ & $322_{-231}$ & $0.982_{-0.008}^{+0.008}$ & $2282_{-133}^{+30}$ & $10_{-0.5}$ & $0.987_{-0.014}^{+0.001}$ & $71.8_{-2.5}^{+0.9}$ & $9.8_{-3.0}$ & $4.21_{-0.12}^{+0.18}$ & $15.92_{-0.28}^{+0.22}$ & $0.0159_{-0.0003}^{+0.003}$ & 85.36/65\\
    \hline
Spec 5 & $2.396_{-0.034}^{+0.19}$ & $0.340_{-0.016}^{+0.017}$ & $27_{-26}^{+81}$ & $0.930_{-0.015}^{+0.020}$ & $2152_{-296}^{+73}$ & $10_{-3}$ & $0.990_{0.020}^{+0.002}$ & $73.4_{-3}^{+1.4}$ & $10_{-5}$ & $4.33_{-0.3}^{+0.26}$ & $16.87^{+1.09}$ & $0.0135_{-0.0010}^{+0.0018}$ & 66.29/65 \\
    \hline
Spec 9 & $2.344_{-0.030}^{+0.018}$ & $0.326_{-0.026}^{+0.032}$ & $36.947_{-10.987}^{+37.320}$ & $0.794_{-0.016}^{+0.018}$ & $2191.660_{-270.021}^{+170.026}$ & $10_{-1.107}$ & $0.982_{-0.013}^{+0.006}$ & $69.495_{-3.577}^{+1.721}$ & $5.003_{-0.400}^{+1.067}$ & $4.246_{-0.208}^{+0.137}$ & $16.861_{-16.861}^{+1.757}$ & $0.0129_{-0.0011}^{+0.0017}$ &  77.73/65\\
    \hline
Spec 10 & $2.398_{-0.032}^{+0.114}$ & $0.349_{-0.022}^{+0.008}$ & $61.094_{-61.048}^{+42.542}$ & $0.917_{-0.012}^{+0.009}$ & $1973.220_{-143.547}^{+39.204}$ & $10_{-0.407}$ & $0.988_{-0.004}^{+0.002}$ & $72.414_{-2.513}^{+1.028}$ & $9.157_{-0.740}$ & $4.300_{-0.172}^{+0.086}$ & $17.006_{-0.607}^{+0.085}$ & $0.0130_{-0.0008}^{+0.005}$ & 59.57/65\\
    \hline
Spec 14 & $2.379_{-0.021}^{+0.022}$ & $0.320_{-0.030}^{+0.011}$ & $74.973_{-74.111}^{+430.321}$ & $0.874_{-0.007}^{+0.008}$ & $1709.270_{-145.151}^{+44.977}$ & $10_{-0.758}$ & $0.972_{-0.014}^{+0.006}$ & $68.098_{-1.471}^{+1.598}$ & $4.966_{-0.130}^{+0.243}$ & $4.310_{-0.079}^{+0.062}$ & $16.980_{-0.436}^{+0.082}$ & $0.0117_{-0.0006}^{+0.008}$ & 96.28/65\\
    \hline
Spec 17& $2.404_{-0.102}^{+0.006}$ & $0.212_{-0.014}^{+0.002}$ & $996.153_{-996.133}$ & $0.902_{-0.004}^{+0.004}$ & $2041.780_{-80.858}^{+45.403}$ & $10_{-0.503}$ & $0.988_{-0.005}^{+0.002}$ & $73.266_{-1.603}^{+1.010}$ & $10_{-1.693}$ & $4.680_{-0.173}$ & $15.993_{-0.523}^{+0.058}$ & $0.0095_{-0.0014}^{+0.07}$ & 99.28/65\\
    \hline
Spec 19& $2.282_{-0.009}^{+0.033}$ & $0.205_{-0.003}^{+0.013}$ & $unconstrained$ & $0.920_{-0.004}^{+0.031}$ & $1372.250_{-252.271}^{+597.106}$ & $10_{-0.270}$ & $0.992_{-0.002}^{+0.001}$ & $73.528_{-1.712}^{+0.440}$ & $8.616_{-1.126}^{+0.576}$ & $4.312_{-0.082}^{+0.034}$ & $18.887_{-0.152}^{+0.101}$ & $0.00970_{-0.00015}^{+0.006}$ & 62.26/65\\    \hline
   \hline
\multicolumn{14}{c}{Joint Fitting Parameter Values}\\
\hline
   \hline
   Spec 4 & $2.513^{+0.005}_{-0.012}$ & $0.368^{+0.007}_{-0.006}$ & $927_{-746}$ & $0.980^{+0.003}_{-0.004}$ & $2265^{+40}_{-42}$ & $9.8_{-0.4}$ & & & & $4.04^{+0.03}_{-0.14}$ & $16.68^{+0.09}_{-0.03}$ & $0.0018^{+0.0007}_{-0.0007}$ &\\  \cline{1-7}\cline{11-13}
   Spec 5 & $2.443^{+0.014}_{-0.012}$ & $0.359^{+0.021}_{-0.03}$ & $47^{+105}_{-11}$ & $0.934^{+0.009}_{-0.009}$ & $2109^{+294}_{-194}$ & $9.9_{-0.5}$ & & & & $4.29^{+0.03}_{-0.6}$ & $16.70^{+0.03}_{-0.08}$ & $0.0014^{+0.0011}_{-0.0005}$ &\\  \cline{1-7}\cline{11-13}
   Spec 9 & $2.291^{+0.007}_{-0.01}$ & $0.387^{+0.008}_{-0.122}$ & $19^{+3}_{-3}$ & $0.847^{+0.003}_{-0.002}$ & $2012^{+58}_{-46}$ & $10_{-0.15}$ & \multirow{0}{*}{$0.987^{+0.003}_{-0.003}$} & \multirow{0}{*}{$71.78^{+0.9}_{-0.3}$} & \multirow{0}{*}{$7.12^{+0.28}_{-0.18}$} & $3.397^{+0.029}_{-0.018}$ & $15.90^{+0.07}_{-0.29}$ & $0.00912^{+0.0004}_{-0.00023}$ & \multirow{0}{*}{$599.66/473$}\\  \cline{1-7}\cline{11-13}
   Spec 10 & $2.418^{+0.012}_{-0.007}$ & $0.347^{+0.007}_{-0.024}$ & $82^{+156}_{-36}$ & $0.916^{+0.002}_{-0.007}$ & $1933^{+47}_{-79}$ & $9.9_{-0.4}$ & & & & $4.26^{+0.08}_{-0.20}$ & $16.878^{+0.116}_{-0.128}$ & $0.00121^{+0.00028}_{-0.00024}$ & \\  \cline{1-7}\cline{11-13}
   Spec 14 & $2.343^{+0.008}_{-0.006}$ & $0.308^{+0.007}_{-0.007}$ & $33^{+4}_{-4}$ & $0.875^{+0.002}_{-0.003}$ & $1743^{+30}_{-30}$ & $9.9_{-0.3}$ & & & & $4.294^{+0.021}_{-0.08}$ & $15.618^{+0.03}_{-0.016}$ & $0.0010^{+0.0001}_{-0.0003}$ & \\  \cline{1-7}\cline{11-13}
   Spec 17& $2.407^{+0.07}_{-0.009}$ & $0.225^{+0.004}_{-0.004}$ & $245_{-133}$ & $0.903^{+0.001}_{-0.001}$ & $2079^{+16}_{-28}$ & $8.7^{+0.26}_{-0.22}$ & & & & $4.335^{+0.022}_{-0.29}$ & $15.690^{+0.07}_{-0.07}$ & $0.00904^{+0.014}_{-0.00013}$ & \\  \cline{1-7}\cline{11-13}
   Spec 19& $2.367^{+0.003}_{-0.007}$ & $0.299^{+0.005}_{-0.013}$ & $999_{-705}$ & $0.9222^{+0.0022}_{-0.0018}$ & $1541^{+23}_{-108}$ & $8.3^{+0.23}_{-0.19}$ & & & & $4.305^{+0.021}_{-0.05}$ & $16.000^{+0.010}_{-0.08}$ & $0.00106_{-0.0008}^{+0.00019}$ & \\ \hline \hline
\end{tabular}}
}
\end{table*}

 \begin{figure*}
\caption{Data to model ratio of the 7 spectra in the G1, Set when fitting with a simple absorbed continuum model: tbabs*(diskbb+powerlaw). Data are rebinned for visual clarity.}
    \includegraphics[width=0.5\textwidth]{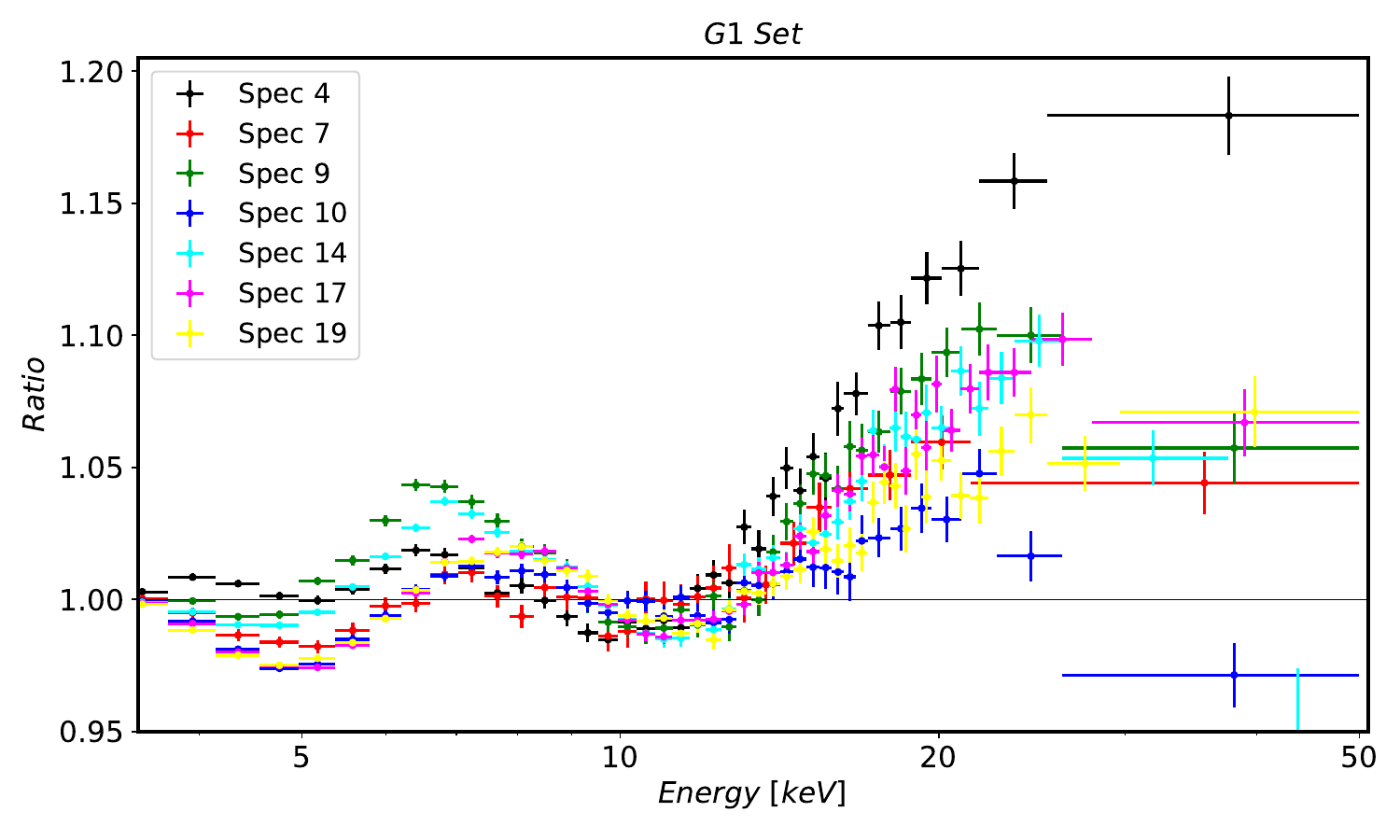}
   \label{fig:figure6a}
\end{figure*}

 \begin{figure*}
\caption{The top panel shows the unfolded spectra obtained from the joint fitting analysis of the G1 Set of spectra with the {\tt relxill} model. The middle and bottom panels show the $\Delta\chi$ when the {\tt relxill} and {\tt relxillD} model formulations were employed to fit the data set, respectively.}
    \includegraphics[width=0.5\textwidth]{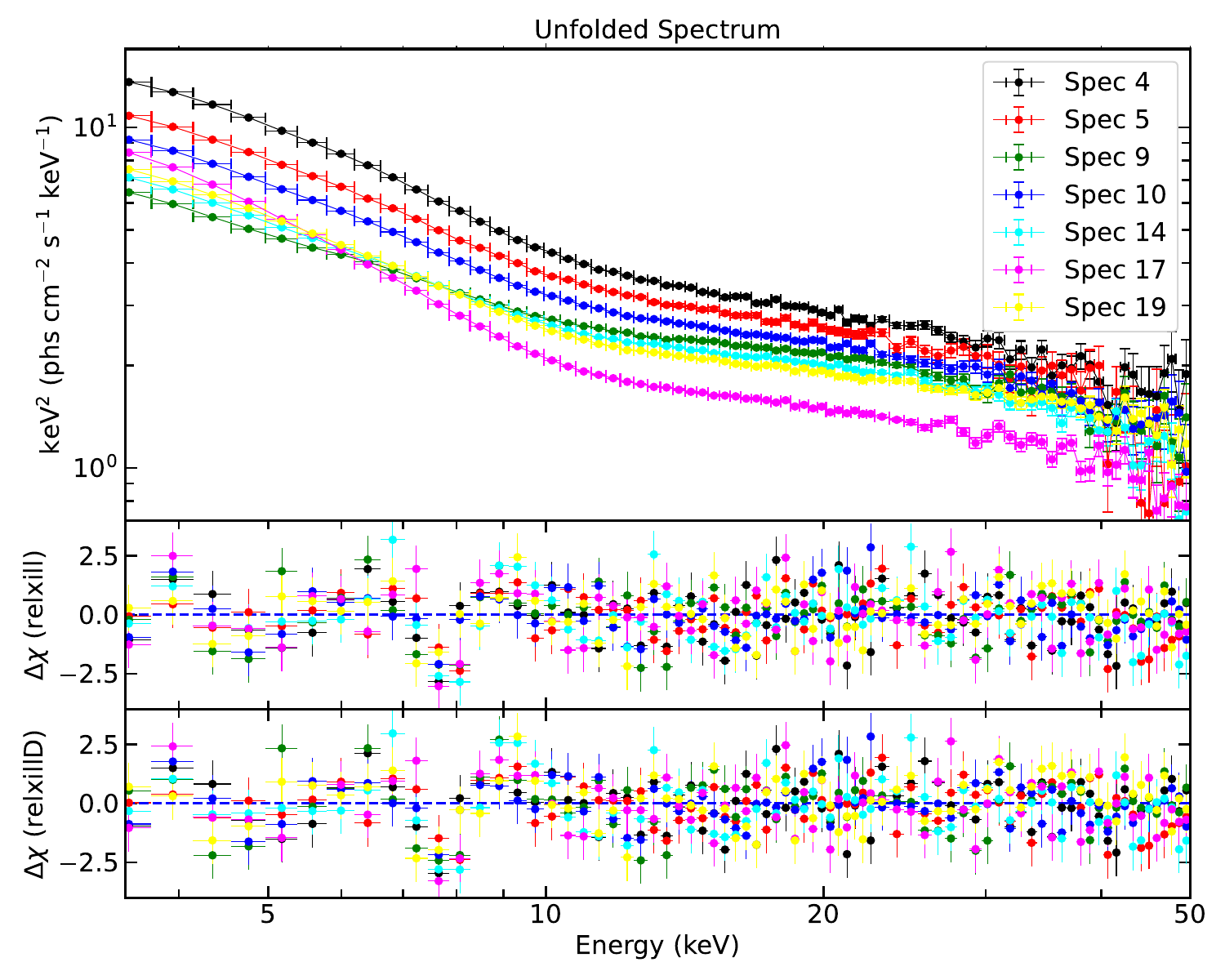}
   \label{fig:figure6}
\end{figure*}

\begin{table*}
\caption{Summary of the best-fit values from the analysis of the G2 Set of spectra of XTE J1859+226 when the emissivity profile of the disk is described by a power law. $^*$ indicates that the parameter is frozen in the fit. The reported uncertainties correspond to the 90\% confidence level for one relevant parameter ($\Delta\chi^2 = 2.71$).}
\renewcommand{\arraystretch}{2}
\setlength{\tabcolsep}{3pt}
\label{table:table5}
\rotatebox{90}{
\begin{tabular}{l|c| c| c| c| c| c | c| c| c | c| c| c }
  \hline
  \hline
 Spec & $\Gamma$ & ${f}_{\mathrm{SC}}$ & $E_{\rm cut}$ & $T_{\rm in}$ & norm & $q$ & $a_*$ & $i$ & $A_{\rm Fe}$ & log$\xi$ & norm & $\chi^2$/$\nu$ \\ 
 & & & [keV] & [keV] & & & & [deg] & & [erg~cm~$s^{-1}$] & & \\\hline
Spec 7 & $2.24_{-0.09}^{+0.03}$ & $0.405_{-0.029}^{+0.024}$ & $79^{15}_{-13}$ & $0.776_{-0.013}^{+0.020}$ & $2560_{-176}^{+182}$ & $10.0_{-1.8}$ & $0.966_{-0.04}^{+0.014}$ & $65.55_{-2.9}^{+2.24}$ & $5.0_{-0.6}$ & $4.00_{-0.09}^{+0.4}$ & $0.042_{-0.006}^{+0.007}$ & 73.97/66 \\
    \hline
Spec 15 & $2.378_{-0.027}^{+0.023}$ & $0.256_{-0.03}^{+0.015}$ & $500_{-213}$ & $0.926_{-0.017}^{+0.013}$ & $2068_{-7}^{+219}$ & $10.0_{-1.3}$ & $0.994_{-0.005}^{+0.003}$ & $76.5_{-4}^{+0.8}$  & $10.0_{-1.2}$ & $3.9_{-0.3}^{+0.3}$ & $0.074_{-0.023}^{+0.010}$ & 62.19/66 \\
    \hline
Spec 21 & $2.385_{-0.025}^{+0.008}$ & $0.122_{-0.026}^{+0.014}$ & $500_{-158}$ & $0.870_{-0.003}^{+0.003}$ & $2060_{-65}^{+51}$ & $1.57_{-0.09}^{+0.23}$ & unconstrained & $71.5_{-2.8}^{+3}$  & $4.20_{-0.6}^{+1.08}$ & $4.30_{-0.06}^{+0.16}$ & $0.026_{-0.008}^{+0.007}$ & 95.8/66 \\
    \hline
Spec 22 & $2.40_{-0.5}^{+0.16}$ & $0.109_{-0.034}^{+0.02}$ & $500_{-263.264}$ & $0.856_{-0.006}^{+0.007}$ & $2250_{-119}^{+285}$ & $8.4_{-2.3}$ & $0.988_{-0.03}$ & $72_{-5}^{+3}$  & $4.2_{-2.3}$ & $3.44_{-0.19}^{+1.03}$ & $0.031_{-0.021}^{+0.025}$ & 42.34/66 \\
    \hline
Spec 23 & $2.24_{-0.05}^{+0.08}$ & $0.053_{-0.031}^{+0.025}$ & $499_{-292}$ & $0.839_{-0.010}^{+0.008}$ & $2354_{-103}^{+135}$ & $10_{-3}$ & $0.994_{-0.011}^{+0.003}$ & $74_{-4}^{+11}$  & $7_{-6}$ & $4.19_{-1.4}^{+0.27}$ & $0.026_{-0.014}^{+0.010}$ & 60.64/66 \\
\hline
   \hline
\multicolumn{13}{c}{Joint Fitting Parameter Values}\\
   \hline
   \hline
Spec 7 &  $2.157_{-0.026}^{+0.03}$ & $0.381_{-0.03}^{+0.021}$ & $63^{11}_{-8}$ & $0.788_{-0.019}^{+0.020}$ & $2475_{-178}^{+90}$ & $10_{-0.4}$ & & & & $4.12_{-0.27}^{+0.27}$ & $0.038_{-0.005}^{+0.004}$ \\ \cline{1-7}\cline{11-12}
Spec 15 & $2.377_{-0.04}^{+0.019}$ & $0.252_{-0.03}^{+0.023}$ & $-500.000_{-225}$ & $0.932_{-0.009}^{+0.011}$ & $1984_{-126}^{+96}$ & $6.4_{-0.7}^{+1.2}$ & \multirow{0}{*}{$0.981_{-0.007}^{+0.006}$} & \multirow{0}{*}{$68.4_{-1.4}^{+1.9}$} & \multirow{0}{*}{$10.0_{-1.7}$} & $4.2_{-0.3}^{+0.4}$ & $0.043_{-0.007}^{+0.008}$ & \multirow{0}{*}{$340.93/342$}\\ \cline{1-7}\cline{11-12}
Spec 21 & $2.27_{-0.03}^{+0.04}$ & $0.100_{-0.03}^{+0.020}$ & $153^{+116}_{-30}$ & $0.858_{-0.005}^{+0.006}$ & $2167_{-97}^{+58}$ & $8.2_{-0.7}^{+0.6}$ & & & & $4.54_{-0.26}$ & $0.026_{-0.007}^{+0.004}$  &  \\\cline{1-7}\cline{11-12}
Spec 22 & $2.159_{-0.228}^{+0.109}$ & $0.089_{-0.025}^{+0.020}$ & $115_{-62}$ & $0.858_{-0.005}^{+0.009}$ & $2215_{-102}^{+59}$ & $7.3_{-0.7}^{+0.8}$ & & & & $3.8_{-0.4}^{+0.6}$ & $0.010_{-0.003}^{+0.006}$ & \\ \cline{1-7}\cline{11-12}
Spec 23 & $2.197_{-0.070}^{+0.050}$ & $0.056_{-0.029}^{+0.021}$ & $373_{-223}$ & $0.848_{-0.007}^{+0.007}$ & $2252_{-94}^{+73}$ & $8.00_{-0.8}^{+1.08}$ & & & & $4.287_{-0.16}^{+0.227}$ & $0.014_{-0.006}^{+0.005}$ &\\ \cline{1-7}\cline{11-12} \hline\hline
\end{tabular}}
\end{table*}

\begin{table*}
\caption{Summary of the best-fit values from the analysis of the G2 Set of spectra of XTE J1859+226, when a power law describes the emissivity profile of the disk and the disk density, is left free. $^*$ indicates that the parameter is frozen in the fit. The reported uncertainties correspond to the 90\% confidence level for one relevant parameter ($\Delta\chi^2 = 2.71$).}
{\scriptsize
\renewcommand{\arraystretch}{2}
\setlength{\tabcolsep}{2.6pt}
\label{table:table6}
\rotatebox{90}{
\begin{tabular}{l| c| c| c| c| c |c| c c c |c |c |c |c}
  \hline
  \hline
   Spec & $\Gamma$ & ${f}_{\mathrm{SC}}$ & kT$_e$ & $T_{\rm in}$ & norm & $q$ & $a_*$ & $i$ & $A_{\rm Fe}$ & log$\xi$ & log N & norm & $\chi^2/\nu$ \\ 
 & & & [keV] & [keV] & & & & [deg] & & [erg cm $s^{-1}$] & & & \\\hline
Spec 7 & $2.305_{-0.017}^{+0.021}$ & $0.410_{-0.022}^{+0.011}$ & $21.94_{-1.24}^{+2.01}$ & $0.774_{-0.011}^{+0.016}$ & $2516_{-431}^{+108}$ & $10_{-1.4}$ & $0.968_{-0.012}^{+0.009}$ & $65.52_{-2.6}^{+2.06}$  & $4.9_{-0.3}^{+1.3}$ & $3.918_{-0.17}^{+0.127}$ & $15.90_{-0.458}^{+0.19}$ & $0.0093_{-0.0003}^{+0.0005}$ & 73.26/65 \\
    \hline
Spec 15 & $2.342_{-0.018}^{+0.015}$ & $0.241_{-0.029}^{+0.007}$ & unconstrained & $0.927_{-0.014}^{+0.009}$ & $2049_{-30}^{+20}$ & $10.0_{-1.3}$ & $0.992_{-0.010}^{+0.001}$ & $76.5_{-3}^{+2.8}$  & $10.0_{-0.9}$ & $3.99_{-1.13}^{+0.29}$ & $16.992_{-0.6}^{+0.124}$ & $0.0077_{-0.0003}^{+0.0005}$ & 66.33/65 \\
    \hline
Spec 21 & $2.341_{-0.010}^{+0.018}$ & $0.109_{-0.018}^{+0.008}$ & $422_{-341}$ & $0.857_{-0.004}^{+0.004}$ & $2181_{-37}^{+81}$ & $10.0_{-0.9}$ & $0.987_{-0.009}^{+0.003}$ & $72.5_{-1.9}^{+1.5}$ & $6.1_{-1.8}^{+1.4}$ & $4.29_{-0.25}^{+0.04}$ & $17.00_{-0.26}^{+0.05}$ & $0.0051_{-0.002}^{+0.0002}$ & 87.15/65 \\
    \hline
Spec 22 & $2.35_{-0.23}^{+0.14}$ & $0.104_{-0.027}^{+0.007}$ & unconstrained & $0.857_{-0.010}^{+0.007}$ & $2236_{-123}^{+103}$ & $8_{-3}$ & $0.988_{-0.012}^{+0.005}$ & $72_{-5}^{+3}$ & $4.645_{-1.029}$ & $3.19_{-0.16}^{+1.04}$ & unconstrained & $0.003_{-0.001}^{+0.002}$ & 42.58/65 \\
    \hline
Spec 23 & $1.88_{-0.05}^{+0.15}$ & $0.032_{-0.015}^{+0.03}$ & $2.8_{-1.5}^{+2.3}$ & $0.837_{-0.021}^{+0.010}$ & $2380_{-91}^{+16}$ & $10_{-5}$ & $0.993_{-0.176}^{+0.001}$ & $73.14_{-11.7}^{+14.2}$ & $5.41_{-0.7}^{+3.18}$ & $3.45_{-0.5}^{+0.09}$ & $18.8_{-2.6}$ & $0.00266_{-0.00004}^{+0.004}$ & 60.07/65 \\
\hline
   \hline
\multicolumn{14}{c}{Joint Fitting Parameter Values}\\
\hline
   \hline
   Spec 7 & $2.284_{-0.014}^{+0.016}$ & $0.418_{-0.013}^{+0.015}$ & $17.92_{-1.02}^{+2.8}$ & $0.787_{-0.016}^{+0.024}$ & $2454_{-165}^{+129}$ & $10_{-0.290}$ & & & & $3.7_{-0.3}^{+0.3}$ & $15.71_{-0.15}^{+0.33}$ & $0.0097_{-0.0004}^{+0.0003}$ &\\  \cline{1-7}\cline{11-13}
   Spec 15 & $2.371_{-0.020}^{+0.004}$ & $0.237_{-0.04}^{+0.014}$ & $1000_{-820}$ & $0.931_{-0.001}^{+0.012}$ & $1943_{-148}^{+65}$ & $6.45_{-0.9}^{+0.18}$ & \multirow{0}{*}{$0.982_{-0.007}^{+0.006}$} & \multirow{0}{*}{$69.1_{-0.8}^{+0.5}$} & \multirow{0}{*}{$5.23_{-0.25}^{+0.5}$} & $4.04_{-0.16}^{+0.24}$ & $16.14_{-0.19}^{+0.08}$ & $0.0067_{-0.0001}^{+0.0001}$ & \multirow{0}{*}{$348.92/337$}\\  \cline{1-7}\cline{11-13}
   Spec 21 & $2.295_{-0.05}^{+0.003}$ & $0.108_{-0.013}^{+0.020}$ & $92_{-45}^{+623}$ & $0.8616_{-0.0003}^{+0.002}$ & $2139_{-42}^{+64}$ & $9.51_{-0.4}^{+0.18}$ & & & & $3.949_{-0.28}^{+0.021}$ & $18.53_{-0.5}^{+0.06}$ & $0.00347_{-0.00002}^{+0.00006}$ & \\  \cline{1-7}\cline{11-13}
   Spec 22 & $2.328_{-0.022}^{+0.005}$ & $0.104_{-0.019}^{+0.002}$ & $1000_{-964}$ & $0.855_{-0.006}^{+0.007}$ & $2258_{-70}^{+70}$ & $7.2_{-0.6}^{+0.3}$ & & & & $3.45_{-1.3}^{+0.29}$ & $16.6_{-0.4}^{+0.4}$ & $0.0025_{-0.0001}^{+0.0002}$ & \\  \cline{1-7}\cline{11-13}
   Spec 23 & $2.234_{-0.05}^{+0.005}$ & $0.059_{-0.001}^{+0.018}$ & $1000_{-956}$ & $0.846_{-0.007}^{+0.007}$ & $2280_{-90}^{+90}$ & $7.952_{-0.9}^{+0.222}$ & & & & $4.020_{-0.23}^{+0.011}$ & $17.46_{-0.4}^{+0.05}$ & $0.00214_{-0.00005}^{+0.0007}$  & \\  \cline{1-7}\cline{11-13}
  \hline \hline
\end{tabular}}
}
\end{table*}

\begin{figure*}
\caption{The top panel shows the unfolded spectra obtained from the joint fitting analysis of the G2 Set of spectra obtained from the {\tt relxill} model. The middle and bottom panels show the $\Delta\chi$ when the {\tt relxill} and {\tt relxillD} model formulations were employed to fit the data set, respectively.}
   \includegraphics[width=0.5\textwidth]{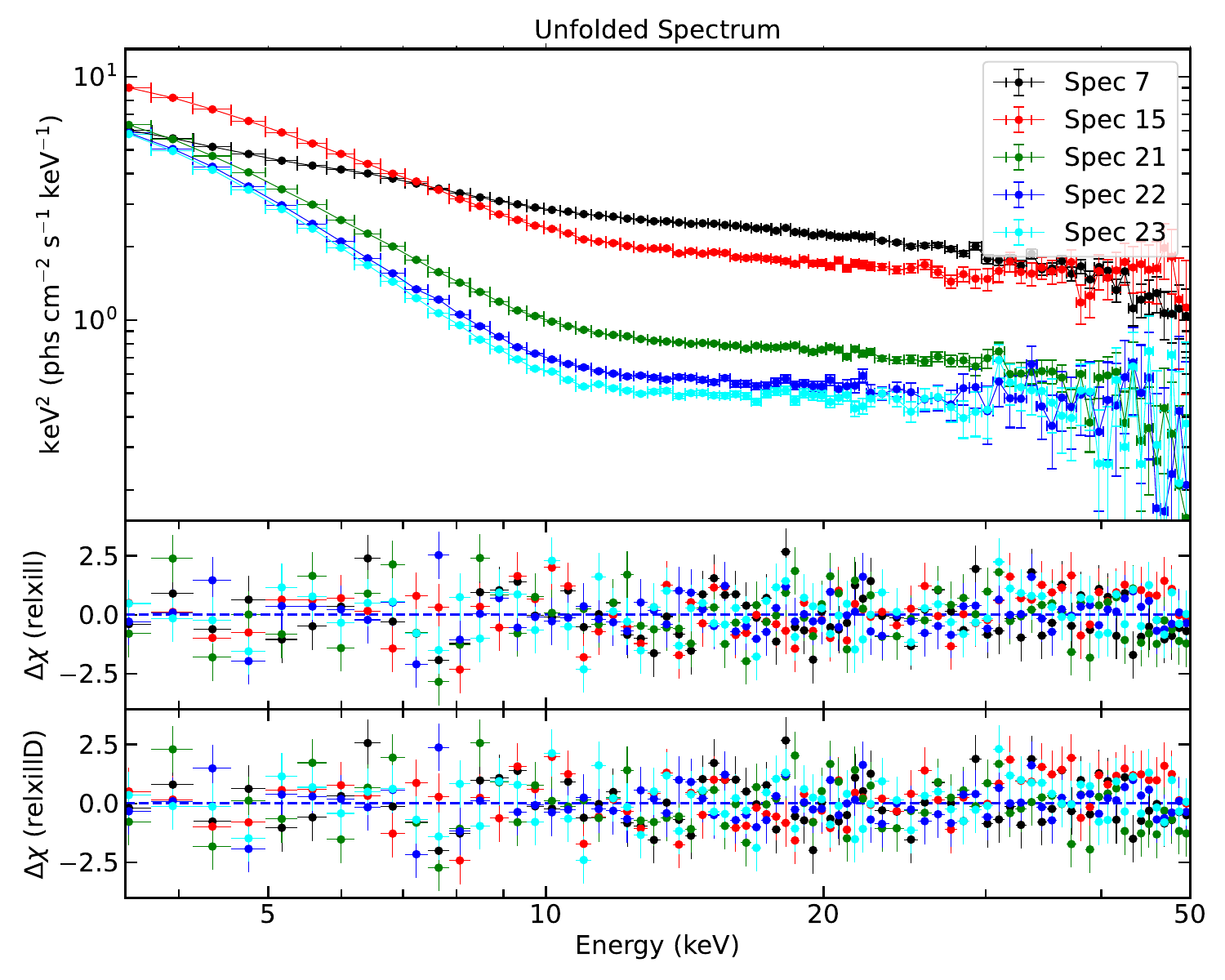}
   \label{fig:figure7}
\end{figure*}


\end{document}